\documentclass[aps,prl,showpacs,twocolumn,amssymb]{revtex4}
\usepackage{amssymb}   
\usepackage{graphicx}
\usepackage{graphics}
\usepackage{color,epsfig}
\usepackage{amsmath}
\usepackage{bm}        
\usepackage{multirow}
\usepackage{booktabs}
\usepackage{pstricks} 
\usepackage[latin1]{inputenc}
\begin{document}

\title{Universality in the cold and ultracold dynamics
                         of the barrierless  D$^{+}$+ H$_2$
                         reaction.}

\author{Manuel Lara\footnote{Corresponding author. E-mail:
 {\em manuel.lara@uam.es}}}
\affiliation{ Departamento de Qu\'imica F\'isica Aplicada, Facultad
de Ciencias, Universidad Aut\'onoma de Madrid, 28049 Madrid, Spain}

\author{P. G. Jambrina}
\affiliation{Departamento de Qu\'imica F\'isica, Facultad
de Qu\'imica, Universidad Complutense, 28040 Madrid,
Spain}

\author{J.-M. Launay} \affiliation{Institut de Physique de Rennes, UMR CNRS
6251, Universit\'e de Rennes I, F-35042 Rennes, France}
\author{F. J. Aoiz\footnote{Corresponding author. E-mail: {\em aoiz@quim.ucm.es}}}
\affiliation{Departamento de Qu\'imica F\'isica, Facultad de Qu\'imica, Universidad Complutense, 28040 Madrid, Spain}

\date{\today}

\begin{abstract}


We have calculated quantum reactive and elastic
cross-sections for the collision  D$^{+}+${\em para}-H$_2$($v$=0, $j$=0)
$\rightarrow$ H$^+$ + HD using the hyperspherical quantum reactive
 scattering method [\textit{Chem. Phys. Lett.}, 1990,  \textbf{169}, 473].
The H$_{3}^{+}$ system is the prototype of barrierless
ion-molecule reactions, apart from its relevance in
astrochemistry. The considered collision energy ranges from
the ultracold regime,  where only one partial wave is open,
up to the Langevin regime, where many of them contribute.
At very low kinetic energies, both an accurate description
of the long-range (LR) region in the potential energy
surface (PES), and long dynamical propagations, up to
distances of 10$^{5}$ a$_0$, are required. Accordingly,
calculations have been carried out on the PES by Velilla
{\em et al.} [\textit{J. Chem. Phys.}, 2008,  \textbf{129},
084307] which accurately reproduces the LR interactions.
Besides, the hyperspherical methodology was recently modified in order
to allow the accurate inclusion of LR interactions while minimizing the computational
expense.
Such implementation is shown particularly suitable for
systems involving ions, where the $\sim R^{-4}$ behavior
largely extends the range of the potential. We find a
reaction rate coefficient which remains almost constant in
a kinetic energy range of more than ten orders of
magnitude. In particular, the value reached in the Wigner
regime,  where only one partial wave is open, is
paradoxically the Langevin classical value (within a few
percent) expected at high energies. Results are discussed
in terms of {\em universality} and related to the recently
published quantum defect theory by Jachymski {\em et al.}
[\textit{Phys. Rev. Lett.}, 2013,  \textbf{110}, 213202].
In this regard, the system has small exothermicity and a
low number of channels, what allows to test such model in a
case where the loss probability at short range is
appreciably far from unity.  The latter parameter has been
rationalized using statistical model assumptions and related
to the products statistical factor.

\end{abstract}

\maketitle

\section{Introduction}
The increasing availability of cold and ultracold samples
of molecules has aroused a great interest in understanding
chemical reactions occurring at very low temperatures.
Given the easiness in the production of (ultra-)cold alkali
metal dimers, attention has mainly focused in reactions
including alkali atoms
\cite{Staanum,Zahzam,Wynar,Mukaiyama,Syassen,hudsonexp}.
This way, in a landmark experiment, reaction rates for
atom+diatom and diatom+diatom collisions involving
fermionic $^{40}$K$^{87}$Rb, at temperatures less than 1
$\mu$K, were recently measured~\cite{Ospel}. In contrast,
advances in the analysis of other type of reactions are
still hampered by  the lack of versatile methods to produce
molecules at low temperatures and the very low densities
achieved. Luckily, emerging experimental techniques promise
to provide detailed information on cold bimolecular
collisions~\cite{Canosa,nuestroFara,Costes,Costes2,FaraBas,Syncro,
Zeem,Nar1,Nar2,Nar3,Nar4}. New experimental approaches
developed by  Narevicius and coworkers
\cite{Nar1,Nar2,Nar3,Nar4}, appear highly promising since
they allow the study of reactions other than those
involving alkalis or possessing permanent electric dipole
moment, while reducing the considered temperatures in more
than three orders of magnitude and hence to ultracold
regime (T$<$1 mK).

With regard to the title system, recent experiments have
been able to stabilize H$_{3}^{+}$  complexes resulting
from the collision  H$^{+}+$ H$_2$ in ion traps for
energies as low as 11K \cite{Gerl}. Experiments to
determine state specific rate coefficients at very low
temperatures are within the reach of current rf ion trap
technology and may be performed in a close future.  The
title reaction may thus provide us with a good chance to
check  and extend our predictive power from the explored
thermal regime, where {\em ab initio} reaction dynamics is
routinely applied, to the low energy regime where such
methodology should fail due to the lack of accuracy. In the
limit, the deep ultracold regime is governed by Wigner
laws~\cite{Sade} and can be described in  terms of  the
scattering length. Paradoxically, the extreme sensitivity
of such parameter to the details of the potential energy
surface (PES), and the action of the surface as a whole,
makes extremely difficult to predict it theoretically  or,
conversely, to deconvolute from its measurement the
underlying interactions. Given the resonant nature of the
scattering length, it largely varies with slight changes
of the interaction potential  whenever a bound state occurs
in the vicinity of the threshold. Only for very light atoms
like metastable He~\cite{Przy,Vassen},  or very particular
atom-atom combinations~\cite{Knoop} it is nowadays possible
to obtain the experimental scattering length  from ``ab
initio'' calculations.
Actually, in the atomic realm the electronic potential
curves are modified "ad hoc" according to the
experimentally measured scattering length in order to
ensure that they are accurate enough to be used in
dynamical studies at low kinetic energies~\cite{Soldan}.

While our knowledge on the behavior of ultracold atomic
systems has impressively increased during the last decades,
it remains as an open question to what extent this might be
achieved for atom+diatom systems, where even the
experimental determination of complex scattering lengths
will have to wait~\cite{Ospel,Jach:02,Jach:03}. {\em Ab
initio} methodology  applied to explore the ultracold
regime in an atom+diatom system like the title system is
not expected to have real predictive power. Instead, in the
spirit of the seminal paper by Gribakin {\em et al.}
\cite{Griba} we will face this study as an effort to
unravel the ``typical'' scattering length that one may
expect. In that reference, the authors obtain a simple
analytical formula which uncovers the way the atomic
interaction  determines the scattering length for a
potential curve (PES in a wide sense) behaving as
$-C_n/R^n$ at LR. Their expression, can be schematically
written as $a=\bar{a} [ 1-\tan(\pi/n-2) \tan (\Phi -
\Phi_0) ] $, where  $\bar{a}$, the {\em mean scattering
length}, does only depend on the LR behavior, and is
proportional to a power of the mass and the $C_n$
parameter. However the semiclassical phase at zero kinetic
energy,  $\Phi$, depends on the shape of the potential at
any $R$, being extremely sensitive to small changes of the
potential when the number of bound levels is large.

In this way, with continuous change of the parameters of
the potential, the phase term, which strongly depends on
the details of the potentials, makes the scattering length
oscillate very quickly about $\bar{a}$, whose dependence on
the potential is weak; the latter thus deserves the name of
of typical or {\em mean scattering length}. In the absence
of bound states close to threshold, $a$ basically coincides
with $\bar{a}$, what leads to the paradoxical idea of {\em
universality}: the result of the collision depends
exclusively on the LR behavior and not on the details of
the strong short-range chemical interactions.

In contrast to the situation in the ultracold regime
commented on above, collisions in the range of $\sim 1$ K
lie within the limits of what can be predicted using the
conventional theoretical tools. Far from being a necessary
stop in our route from the thermal to the ultracold regime,
they have an interest on their own. Apart from being of
astrophysical relevance, they report the combined influence
of LR and short-range (SR) interactions. While the
ultracold scenario favors LR interactions, leading to
universality in  some extreme cases\cite{Fara}, the thermal
regime  and its higher kinetic energies makes SR chemical
forces prevail. In the cold regime both SR and LR
interactions play the game and the use of a balanced PES
which describes accurately the whole configuration space is
required. Sometimes, their roles may be independent enough
to allow to associate trends in the behavior to  different
regions of the PES, offering insight in the underlying
dynamics \cite{Lara2}.

Focusing on the title system, reactions of ions with
neutrals do not commonly have a significant barrier and,
due to their LR attractive potentials ($n=4$), exhibit
large cross sections. This renders them especially relevant
at the low temperatures typical of the interstellar medium
(see refs. \cite{S:CR92,H:CSR01,SB:ARAC08} and references
therein).   In the past decades, much experimental effort
has been dedicated to extend the temperature range down to
a few K. The difficulties associated with the handling of
small  relative translational energies in ion--molecule
reactions have been overcome through the use of supersonic
jets \cite{RM:IJMSIP87,HMRSZS:IJMSIP90,S:IJPC98,SR:ACR00},
guided and merged beams, and ion traps
\cite{TG:CP74,G:ACP92,T:CR92,G:JCSFAR93,G:PS95}. In
particular, special attention has been paid to the study of
hydrogen-like ions and specifically to the investigation of
the H$_{3}^{+}$ system.

The H$_{3}^{+}$ molecule is the most abundant triatomic
ionic species in dense interstellar clouds \cite{MGHO:AJ99}
and in many cold hydrogen plasmas \cite{HP:PP02,
MGHT:JPCA06}. This ion is also formed as a strongly bound
intermediate in collisions of H$^{+}$ with H$_{2}$.  Due to
its apparent simplicity, the system H$_{3}^{+}$
constitutes a prototype in the field of ion-molecule
reactions. As a result it has been a favorite of both
theoretical and experimental studies. Depending on the
total energy, collisions of protons with hydrogen molecules
can have different outcomes, including rovibrational energy
transfer, charge transfer, dissociation of the molecule
into its atomic components, as well as radiative
association leading to stable H$_3^+$. At energies below
$\approx$ 1.7 eV proton exchange is the only reactive
channel and we will limit our attention to this particular
process. Early calculations starting in the seventies
\cite{KPWT:JCP74,GNST:CP80,G:SASP82,SV:CP87,BS:JCP94}
disclosed the main characteristics of the reaction
dynamics. It was seen to evolve from a low energy behavior,
dominated by capture into a strongly interacting complex,
followed by a statistical breakdown of the three-atom
complex, to the appearance of dynamical constraints with
growing energy, caused by increasingly direct collisions
with shorter interaction times. These short interaction
times do not allow for a complete randomization of the
energy, angular momentum, and nuclear scrambling within the
reaction intermediate. These calculations, based on simple
statistical models, semiempirical PESs, and a limited
number of classical trajectories, were able to account
reasonably well for the available experimental values of
cross sections and rate coefficients
\cite{OT:JCP74,FABFGF:JCP74,HAS:JCP81,VHL:JCP82,M:Diplom,G:ACP92},
although not without considerable discrepancies.

Over the last two decades, great progress was achieved in
the construction of accurate potential surfaces for the
H$_{3}^{+}$ system
\cite{IY:JCP95,ARTSP:JCP00,KBNN:JCP02,KJ06,VAV:JCP07,VLABP:JCP08,BCJK:JCP09}
and in the development of gradually more rigorous
theoretical approaches for the investigation of the nuclear
motion. Refined statistical treatments, exhaustive
quasiclassical trajectory (QCT) calculations, and time
independent (TI) as well as time dependent wave-packet
(TDWP) quantum mechanical (QM) methods of varying accuracy
were applied to the study of the H$^{+}$+H$_{2}$ reaction
dynamics (see for instance Refs.
\onlinecite{TKI:JCP00,GAPR:JCP05,LCH:JPCA05,ASGM:JCP07,AGS:JCP07,CGRHLBABTW:JCP08,ZZC:JTCC09,ZC:JTCC09,GHJAL:JCP09,ZRGRASG:JPCA09,JAEHS:JCP09,ZLXC:CP10,JABSBH:PCCP10,JAAHS:PCCP10,GM:JPCA11}
and references therein). Many of these theoretical works
were centered on methodological aspects, emphasizing the
comparison between different approaches. As a result of
these studies it became clear, that despite the apparent
simplicity of the reaction considered, all theoretical
methods met with problems for the description of its
dynamics.

Of all the possible isotopic variants of the H$^{+}$ +
H$_{2}$ reaction, that of the deuteron with H$_2$ is
exothermic due to the different zero-point energies (ZPE)
of reactants and products and thus it is appropriate for a
study at the cold and ultra-cold energy regimes. The
isotopic substitution allows for a straightforward
identification of reactants and products using methods of
mass spectrometry. In addition, D$^{+}$ + H$_{2}$ can play
an important role in the unusual deuterium fractionation
observed in many cold space environments
\cite{W:AJ73a,GS:PSS02,M:AG05}. In fact, isotope selective
effects due to deuterated variants of the H$^{+}$+H$_{2}$
reaction are observable even in room temperature discharges
of H$_{2}$/D$_{2}$ mixtures \cite{JCHT:PCCP11}.

In the present work, we have calculated quantum reactive
and elastic cross-sections and rate coefficients for the
collision D$^{+}$ + {\em para}-H$_2$ $\rightarrow$ H$^+$ +
HD using the hyperspherical quantum reactive scattering
method \cite{Launayfirst}. We have considered collision
energies ranging from the ultracold regime, where only one
partial wave is open, to the Langevin regime, where many of
them contribute. Following the classical and venerable
Langevin model, it is expected that barrierless ion-neutral
reactions, associated to  potentials varying as $R^{-4}$ at
LR, will display a constant reaction rate coefficient at
high energies (see below). At the same time, the Wigner
threshold laws predict also a constant value at very low
collision energies. However, the latter does not state
whether its value would be high or null, or if it coincides
with the Langevin high energy limit. The way both rates
(for high and low energy) are related was to be determined.
As it will be shown, they basically coincide for the title
system, leading to a nearly constant rate coefficient  in
the whole considered energy range.

At very low  kinetic energies, LR interactions (usually
disregarded in both PESs and dynamical calculations) are
essential, as they determine the amount of incoming flux
which reaches the short range region where rearrangement
may occur. Both accurate descriptions of the LR region in
the PES, and long dynamical propagations (up to distances
of hundreds of thousands of a.u.) are thus required. Among
the existing PESs for the title system, that by Velilla
{\em al.} satisfies the first requirement since it
accurately reproduces the LR interactions
\cite{VLABP:JCP08}.  The second requirement is fully
satisfied by the hyperspherical methodology, recently
modified to allow the accurate inclusion of LR interactions
while minimizing the computational expense~\cite{Lara2}.

In two recent works, the dynamics of the title system was
analyzed for energies above the cold
regime~\cite{Honverr,Honv1,Honv2} using the same PES. The
hyperspherical quantum reactive scattering
method~\cite{Launayfirst} was also used although with an
implementation that is usually employed in the study of
thermal and hyperthermal reaction.  In their approach, the
propagation is also carried out in hyperspherical
coordinates but up to a hyperradius which is large enough
for the potential energy to be negligible in comparison
with the collision energy.  The use of such implementation
would be infeasible in the energy regime considered in this
work.  In fact, the authors of Ref.~\cite{Honverr} had
problems to converge the lowest energies they considered
(on the order of 10 K) and were forced to extend the
propagation up to $\rho$=40 a.u. As we will show, small
methodological changes in the hyperspherical method allow
the convergence of calculations for much lower kinetic
energies at a reasonable computational
expense~\cite{Lara2}. These changes are used for the first
time in the present work in order to converge scattering
results corresponding to energies as low as $10^{-8}$K for
a system where the $\sim R^{-4}$ behavior largely extends
the range of the potential.

The paper is structured as follows. In the next section, we
will briefly describe the theoretical methodology,
recalling the hyperspherical approach and its recent
improvements, and providing details on the considered PES
and the calculation of effective potentials. The results
from the dynamical calculations will be shown and discussed
in section III. Finally, a summary of the work and the
conclusions will be given in Section IV.

\begin{figure}[t]
 \begin{center}
\includegraphics[width=60ex]{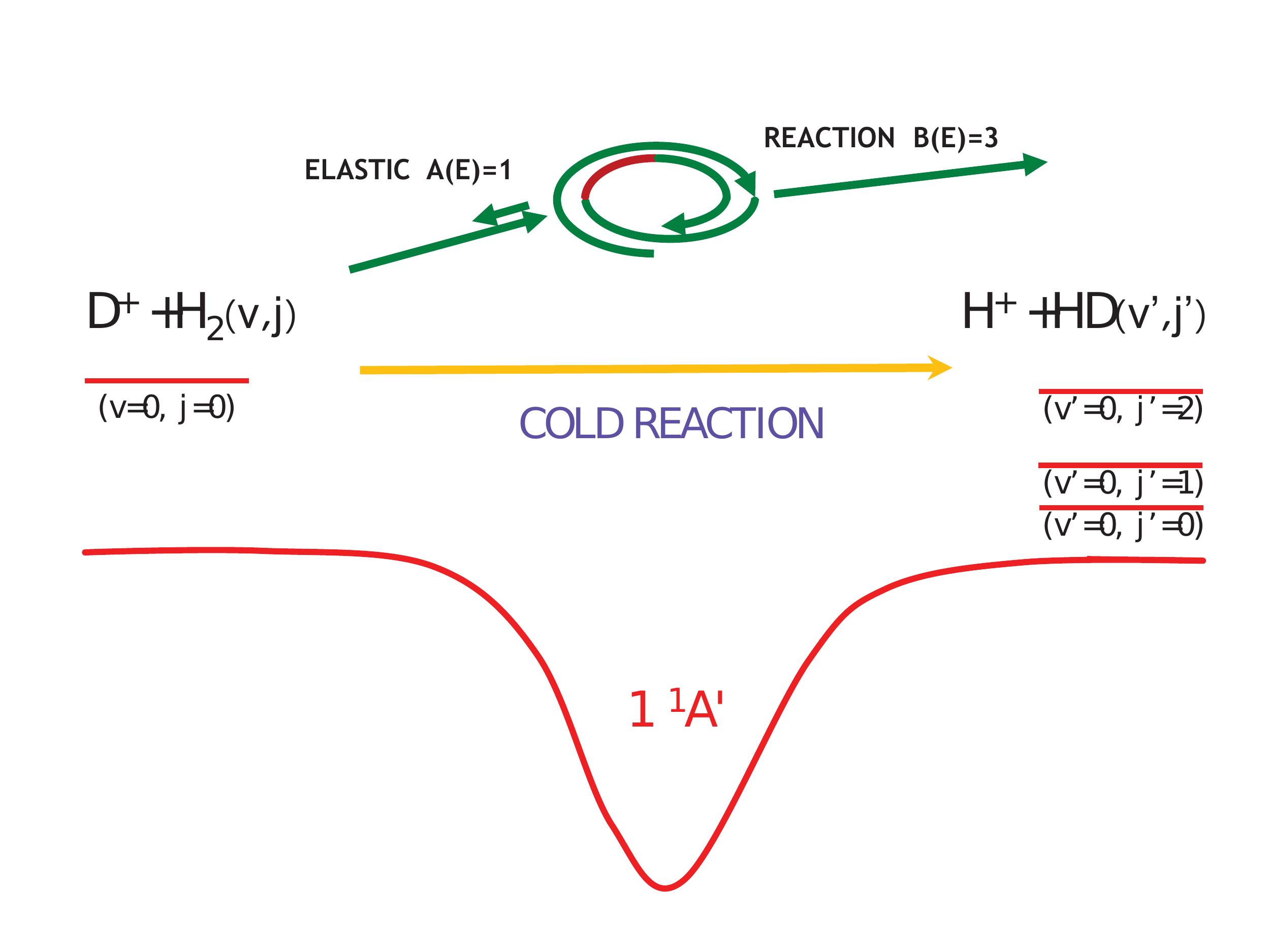}
\caption{Representation of the minimum energy path and the rovibrational
states involved in the reactive collision on the ground PES.
The values of $A(E)$ and $B(E)$ for the collision with $J=0$ are also
shown (see the text for definitions).} \label{fig0}
\end{center}
\end{figure}

\section{Theoretical Methods}
\label{Themet}

Not so long ago, errors in {\em ab initio} PES were in the
order of 1 kcal/mol. Hence it was indeed pointless to
attempt an accurate description of the much smaller
interactions at LR. Moreover, the high kinetic energies
usually considered made possible to assume free-evolution
in the dynamics from distances as short as 20-30 a$_0$.
This is a general practice in both time-dependent and
time-independent dynamical methods in order to avoid very
long-propagations. However, as we will show, in the
unexplored realm of processes at extremely low
temperatures, LR interactions cannot be neglected and
theoretical tools (used in both electronic structure and
nuclear dynamics  methodologies) have to be taken to their
limit.

\subsection{Dynamical methodology}

The hyperspherical quantum reactive scattering method
developed by Launay {\em et al.}~\cite{Launayfirst} has
been widely used and checked in the context of thermal
reactive scattering \cite{hon04}. Already applied to
describe ultracold collisions of
alkalis~\cite{Sol02,Quem04}, the method has proven its
capability in reactive processes at very low temperatures,
wherein time-dependent techniques (of common use at thermal
energies) face insurmountable handicaps. Besides, the
particular implementation it uses (in contrast with others
involving hyperspherical coordinates), makes it very
suitable for systems whose PES shows very deep wells, as it
is the case for the title system.

Recent modifications of the method, performed in order to
allow the accurate inclusion of small anisotropic LR
interactions in the dynamics, were described in depth in
ref.~\cite{Lara2}. In what follows, we will simply recall
the basic concepts, referring to previous works for more
details.

In the hyperspherical quantum reactive scattering method,
developed by Launay~\cite{Launayfirst}, the configuration
space is divided into inner and outer regions. The
positions of the nuclei in the inner region are described
in terms of hyperspherical democratic coordinates. The
logarithmic derivative of the wavefunction is propagated
outwards on a single adiabatic PES. At a large enough value
of the hyper-radius the logarithmic derivative is matched
to a set of suitable functions, called asymptotic
functions, which provide the collision boundary conditions,
to yield the scattering S-matrix. When working at thermal
energies, these functions are the familiar regular and
irregular radial Bessel functions. They account for the
presence of a centrifugal potential without the need of
extending the calculation to very large distances, where
the centrifugal potential vanishes.

Small albeit important methodological improvements were
recently introduced~\cite{Lara2} for these functions to
account also for anisotropic LR potentials, thus enabling
the study of cold and ultracold collisions. These
improvements involve the definition of numerical asymptotic
radial wavefunctions which are adapted to a specific LR
behavior in an analogous way as radial Bessel functions are
adapted to the presence of a centrifugal potential. These
asymptotic radial functions thus provide the collisional
boundary conditions in the presence of general potentials
at LR. They are obtained by solving a system of radial
differential equations in Jacobi coordinates~\cite{Lara2}.
Although they imply long propagations from the
``asymptotic'' region up to the matching hyper-radius
(chosen in our case at 35 a.u.) the expense is minimal:
the number of coupled equations is reduced to one for
$j=0$. In this way, we can account for the potential at LR
while avoiding propagations in hyperspherical coordinates
up to extremely large intermolecular separations. Let us
note that to converge the elastic cross-sections for the
lower partial waves at a collision energy of $10^{-8}K$,
distances on the order of hundreds of thousands of a$_0$
had to be considered. This would be totally unfeasible
considering only a propagation in hyperspherical
coordinates.

\subsection{The electronic energies}
For the H$_3^+$ system, the only process that can take
place below 1.7 eV is the non-charge-transfer proton
exchange process, and the reaction can be rigorously
described only using the ground adiabatic PES. A deep well
($\approx 4.5eV$) and a small exoergicity (that of the
difference of the zero point energies) are the main
features of the PES. A diagram of the minimum energy path
and the main energetics of the process is shown in
Fig.~\ref{fig0}.

New experimental achievements and improvements in the
computational performance of electronic calculations call
nowadays for more than spectroscopic accuracy, with the aim
of building PESs able to describe the whole configuration
space (both LR and SR regions) in a balanced way. Such
demand is compulsory when studying cold collisions. The PES
by Velilla {\em et al.}~\cite{VLABP:JCP08}, chosen to study
the dynamics of the title system, seems to fulfill these
requirements. It accurately reproduces the LR interactions,
which are included in the functional form of the potential.
The asymptotic form of the PES in reactant Jacobi
coordinates $(r,R,\theta)$, is given by:
\begin{eqnarray}
\label{eq:LR}
V_{\rm LR}(r,R,\theta)=Q_2(r) P_2 ( \cos \theta ) R^{-3}- [1/2 \alpha_0(r) \nonumber \\
+ 1/3(\alpha_{\parallel}(r)-\alpha_{\perp}(r)) P_2 (\cos\theta ) ] R^{-4} + \ldots
\end{eqnarray}
where $Q_2(r)$ is the quadrupole moment, and
$\alpha_{0}(r)$, $\alpha_{\parallel}(r)$ and
$\alpha_{\perp}(r)$ are, respectively, the average,
parallel and perpendicular polarizabilities of H$_2$. In
spite of their dependence on $r$, their values at the
equilibrium distance of H$_2$ are~\cite{VLABP:JCP08}: $Q_2
\sim 0.5 a.u.$, $\alpha_0 \sim 5.4  a.u.$,
$\alpha_{\parallel} \sim 6.8 a.u.$ and $\alpha_{\perp}
\sim 4.8 a.u.$ As the expression shows, the dominant
contributions involve the charge-quadrupole interaction,
varying as  $\sim R^{-3}$ and the charge-induced dipole,
varying as $\sim R^{-4}$. The surface implements a recent
method, based on the use of a change of variables in LR
terms~\cite{VLABP:JCP08}, to solve problems in fittings
associated to the divergence at SR of typical asymptotic
terms ($\sim R^{-n}$). An expansion in analytical LR terms
was considered {\em a priori} in the fit, and it was the
difference between the energies given by such an expansion
and the {\em ab initio} energies what was fitted using
conventional techniques~\cite{Agua}.

In order to show the significance of LR effects, the
results obtained by using the PES by Velilla {\em et al.}
will be compared with the ones obtained using its
``ancestor'': the surface by Aguado {\em et
al.}~\cite{ARTSP:JCP00}. It relies on 8469 configuration
interaction {\em ab initio} points and a global fit using
diatomics-in-molecules approach together with three body
corrections. The PES by Velilla {\em et al.} is a recent
refinement of the PES by Aguado {\em et
al.}~\cite{ARTSP:JCP00}, being based on the same set of
{\em ab initio} points; however it pays especial care in
the fitting of the LR region.

The accuracy of the PES, in absolute energies, is in the
range of 0.1 cm$^{-1}$ (0.14 K) in the interaction region
and less than 1 cm$^{-1}$ (1.4 K) in the LR region.
However, the error in relative energies (differences) from
one point in the configuration space to another might well
be less than the tenth of such~\footnote[7]{Private
communication by the authors of Ref.~\cite{VLABP:JCP08}}.

\subsection{The effective potentials}
Let us label with ${\bm l}$ the orbital angular momentum of
the atom with respect to the center of mass of the diatom,
and with ${\bm j}$ the rotational angular momentum of the
latter. The total angular momentum of the nuclei (conserved
in an adiabatic approach) is given by
$\bm{J}=\bm{j}+\bm{l}$. A convenient basis in order to
expand the nuclear wavefunction in the LR region is the one
characterized by quantum numbers $(J,M,v,j,l)$, represented
as $\varphi^{J M}_{v j l }$, with ($v,j$) the rovibrational
quantum numbers of the diatom, $l$ the relative orbital
angular momentum and ($J,M$) the total angular momentum
and its projection on the Space-Fixed (SF) $Z$ axis. Let us
consider the matrix of the electronic potential expressed
in this basis, $\langle \varphi^{J M}_{v j l} | V  |
\varphi^{J M}_{v' j' l'} \rangle$. Its diagonal elements
are useful in order to understand the dynamics because the
diabatic effective potential felt by the colliding partners
at a distance $R$ when approaching in the state $\varphi^{J
M}_{v_0 j_0 l_0}$ is given by $\langle \varphi^{J M}_{v_0
j_0 l_0} | V  | \varphi^{J M}_{v_0 j_0 l_0} \rangle (R)+
l_0(l_0+1) \hbar^2 /2 \mu R^2 $. For simplicity reasons, we
will design with $\langle  V \rangle_{l,l'}(R)$ the matrix
element  $\langle \varphi^{J M}_{v j l } | V  | \varphi^{J
M}_{v j l'} \rangle$.

Regarding the calculation of the potential matrix $\langle
V \rangle_{ l, l'} (R)$, it is convenient to calculate
first the one associated to another basis, labeled by the
projection $\Omega_j$ of ${\bm J}$ on the Body--Fixed (BF)
coordinate system, whose $z$-axis is chosen along the
reactant Jacobi $\bm R$ vector. This BF basis set is given
by
\begin{eqnarray}\label{eq:jotas}
\phi^{J M}_{v  j  \Omega_j }=\frac{\chi_{v,j}(r)}{r}
\sqrt{\frac{2J+1}{4\pi}}D_{M \Omega_j}^{J*}(\alpha,\beta,\gamma)Y_{j
  \Omega_j}(\theta,0),
\end{eqnarray}
where $\chi_{v,j}(r)$ is the radial rovibrational wave
function, $Y_{j \Omega_j}(\theta,\phi)$ the spherical
harmonics, and $D_{M \Omega_j}^{J*}$ denotes a Wigner
rotation matrix element with $(\alpha,\beta,\gamma)$ being
the Euler angles corresponding to the transformation
between SF and BF frames. The matrix elements of the PES
$V(R,r,\theta)$ in this basis are given by
\begin{widetext}
\begin{equation}\label{clarito}
\langle  V \rangle_{\Omega_j, \Omega_j'}(R)= \delta_{\Omega_j  \Omega_j'} 2 \pi  \int  \chi_{v,j}^2(r)
 Y_{j \Omega_j}^2(\theta,0) V(R,r,\theta) \sin\theta\; dr\; d\theta
\end{equation}
Once the potential matrix is calculated in the BF
frame, we change to the SF basis (what involves
a combination using $3j$ symbols) thus obtaining:
\begin{eqnarray}\label{camio}
\langle  V \rangle_{l,l'}(R) & = & (-1)^{l+l'} \sqrt{2l+1}\sqrt{2l'+1} \\
&&\times  \sum_{\Omega_j} \left(\begin{array}{ccc} j & l & J  \\ \Omega_j & 0  & -\Omega_j \end{array}\right)
\left(\begin{array}{ccc} j & l' & J  \\ \Omega_j & 0  & -\Omega_j \end{array}\right)  \langle  V \rangle_{\Omega_j, \Omega_j}(R). \nonumber
\end{eqnarray}
\noindent
\end{widetext}

\section{Results and Discussion}

\subsection{The reaction at low collision energy: the Numerical-Capture Statistical model}

The classical Langevin capture model for a potential
behaving as $R^{-4}$ is usually applied to rationalize the
cross-sections and rate constants corresponding to an ion +
neutral collision. Even when the asymptotic expression in
Eq.~\ref{eq:LR} contains a charge-quadrupole term $\sim
R^{-3}$, collisions with $j=0$ correspond to the case $n=4$
for this reactive system. Indeed, the effective potential
which governs the collision behaves as $-C_4/R^4$: the
integral $\langle j=0 | P_2 | j=0 \rangle$ is null because
$ |j=0 \rangle$ is equivalent to the Legendre polynomial
$P_0$, and the contribution of the term $\sim R^{-3}$ (as
well as the anisotropic polarization term) in
Eq.~\ref{eq:LR} vanishes. Let us remark that, in contrast
to neutral + neutral reactions, the  $\sim R^{-4}$ behavior
extends the range of the interaction to very large
distances.

When applied to the title system the Langevin model (used
for collisions dominated by attractive potentials and high
reaction probabilities) it becomes equivalent to assume
that the system is captured in the complex and leads to
reaction whenever the collision energy overcomes the
centrifugal barrier. In fact, starting from the expression
$\sigma (E)=(\pi / k^2) \sum (2l+1)P_r(E)$ and:  i)
calculating the centrifugal barrier height for each $l$
corresponding to an analytical potential $-C_4/R^4$; ii)
assuming that $P_r(E)$ is zero below the centrifugal
barrier, and the unity over it; and iii) replacing the
summation with an integral over impact parameters, one gets
the well known Langevin expression for the cross-section
$\sigma_{\rm L}(E)= 2 \pi (C_4 /E)^{1/2}$ (or $k_{\rm
L}(E)= 2 \pi (2 C_4 / \mu)^{1/2}$ for the rate coefficient,
where $ k_{\rm L}(E)=v_{\rm rel} \sigma_{\rm L}$ and
$v_{\rm rel}$ the relative velocity). The value of $C_4$
corresponding to the polarization potential  can be easily
estimated starting from $\alpha$, the average
polarizability of the molecule, and $q$, the charge of the
ion, by using the expression $C_4=\alpha q^2 /2$. Taking
$\alpha$=5.41 a.u. from the literature~\cite{alpha}, we get
a value of $C_4$=2.70 a.u.. More accurately, one should
consider the $C_4$ associated to the PES and calculated
from the effective potential corresponding to $J=0$: as
expected, the latter behaves asymptotically as $-C_4/R^{4}$
with a value for $C_4$ given by 2.71 a$_0$.

\begin{figure}[ht!]
 \begin{center}
\includegraphics[width=55ex]{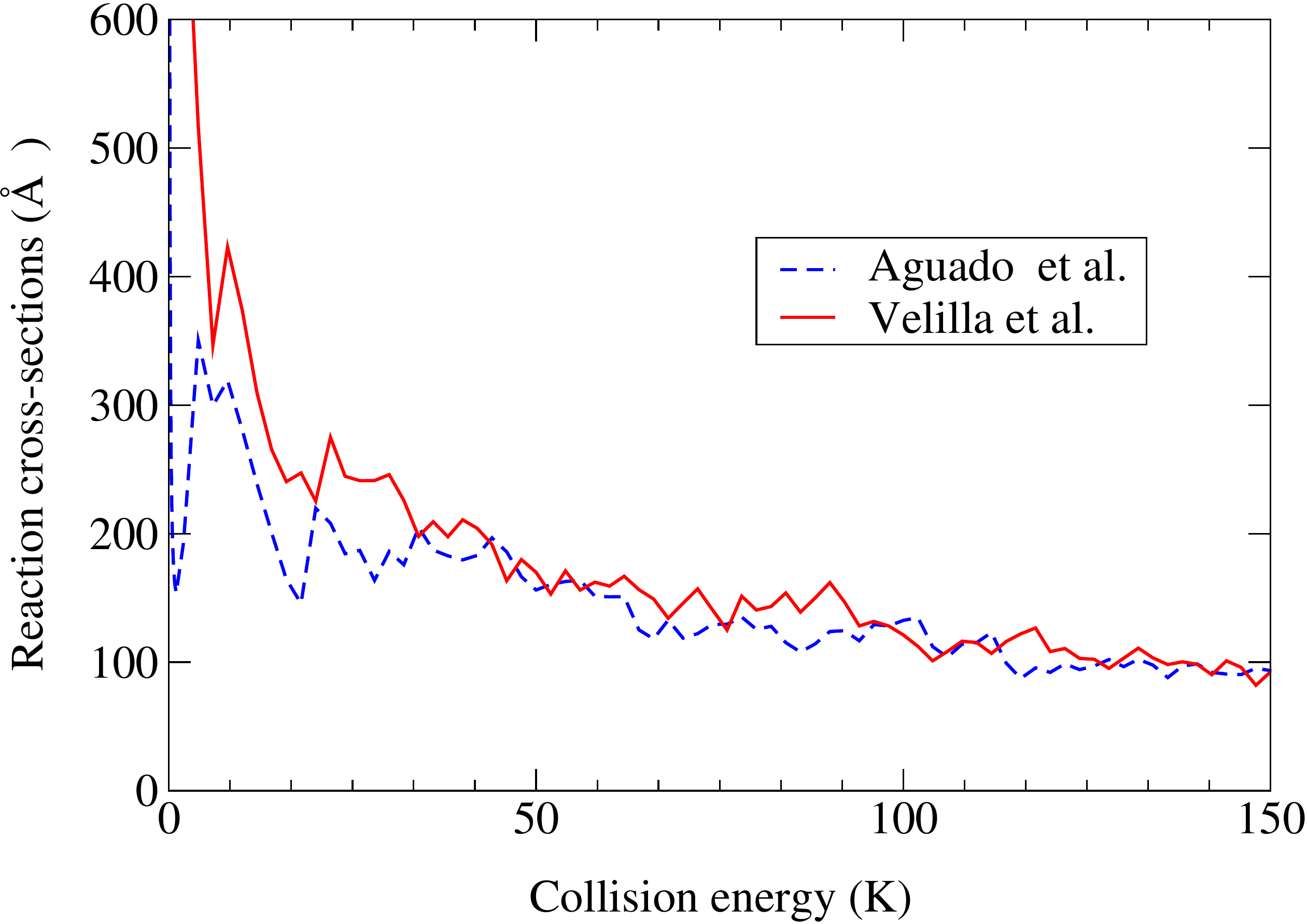}
\caption{Reaction cross-sections for the collision
D$^{+}$+H$_2$ ($v$=0, $j$=0)  at low
collision energies. Results for both surfaces are compared. The surface by
 Aguado {\em et al.}  seems somewhat less reactive than the one by Velilla {\em et al.}
} \label{fig1}
\end{center}
\end{figure}

\begin{figure}[h]
\begin{center}
\includegraphics[width=60ex]{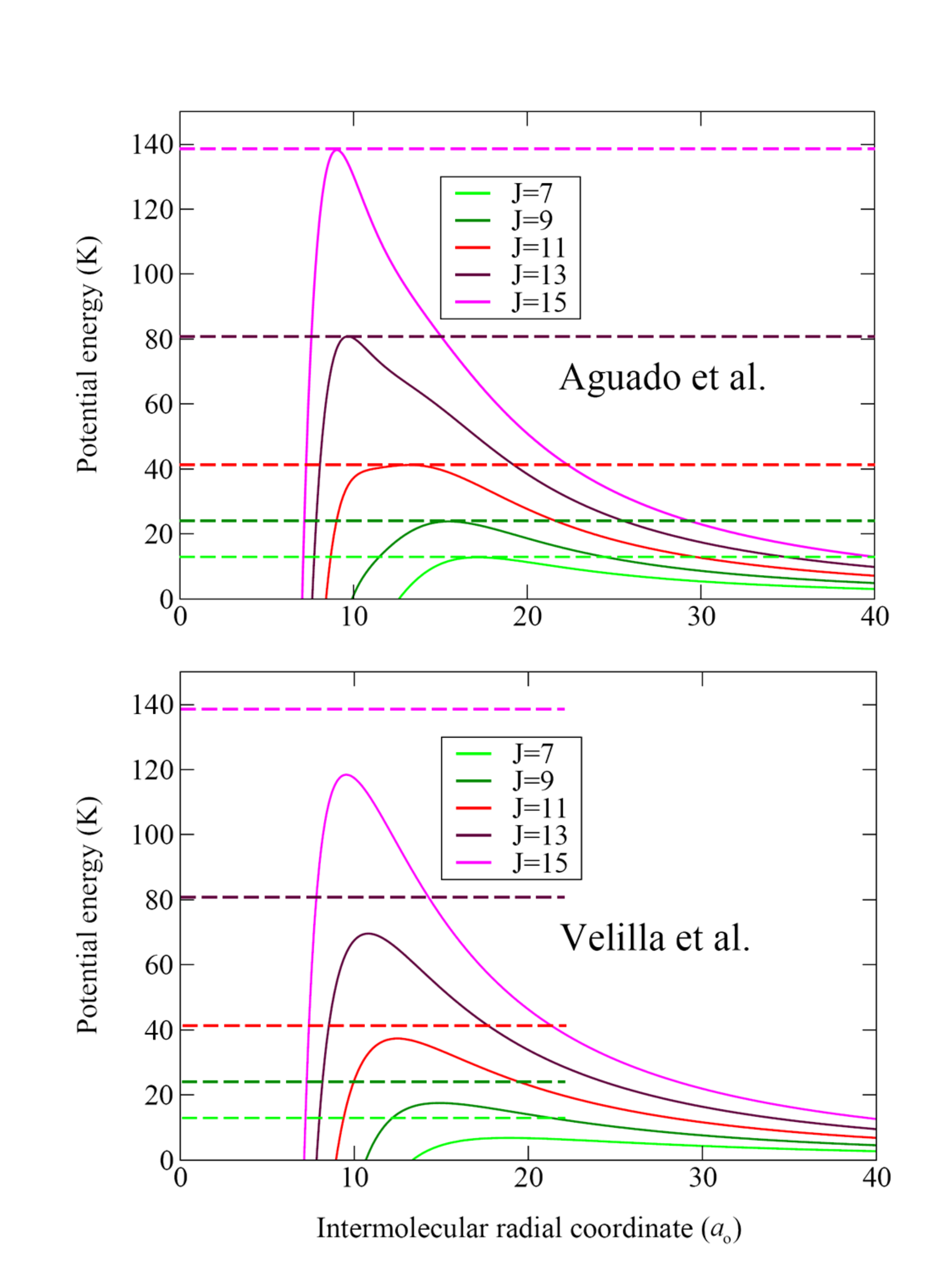}
\caption{The effective potentials calculated for the both PES considered.
The centrifugal barriers are lower in the case of the surface by Aguado
 {\em et al.} what explains the higher reactivity observed at low energies.
} \label{fig2}
\end{center}
\end{figure}
The reaction cross--sections for low collision energies
obtained using both considered surfaces are shown in
Fig.~\ref{fig1}. The PES by Aguado {\em et al.} seems to be
somewhat less reactive than the one by Velilla {\em et
al.}. In order to understand such difference, we have
calculated the effective potentials associated to low
partial waves and they are shown in Fig.~\ref{fig2}. As can
be seen, the centrifugal barriers are lower for the surface
by Velilla {\em et al.} what explains the higher
reactivity. The polarization term, accurately included in
the PES, pulls the centrifugal barriers down, thus allowing
more incoming flux to reach the inner region and to lead to
reaction. Once we have shown the significance of the LR
interactions, even at energies well above the cold energy
range, we will restrict our calculations to the PES by
Velilla {\em et al.}

The same values from Fig.~\ref{fig1} are plotted in
logarithmic scale in the high energy region of
Fig.~\ref{fig4}, where the results for much lower collision
energies are also shown together with the Langevin
predictions. Rate coefficients are shown in Fig.~\ref{fig5}
and compared with $k_{\rm L}(E)$. First of all, let us
focus at high collision energies, where many partial waves
are open and Langevin model could be valid. Our theoretical
cross-sections are found smaller than the Langevin
prediction and this is not surprising. Actually, Langevin
model is a crude approximation. We can try to improve it
in, at least, three ways:  i) calculating  the heights of
the centrifugal barriers numerically, using the effective
potentials, thus accounting for deviations from $R^{-4}$ at
short-range; ii) performing the summation over partial
waves instead of the integral, what will introduce sudden
'steps' in $\sigma(E)$; iii) correcting with a statistical
factor (6/7 in this case, as discussed hereunder) to
account for the probability of a complex to decompose into
the products arrangement channel. We will call this
``improved'' Langevin model as 'numerical-capture
statistical model'. The results of the refined model are
shown in the high energy region of Figs.~\ref{fig4} and
\ref{fig5}. The predicted cross-sections and rates are
lower, in better agreement with the calculations. If
desired, the steps in the model can be easily suppressed
by approximating each centrifugal barrier for an analytical
barrier, and using the associated transmission probability
instead of the step function.

The factor 6/7 deserves more detailed comments. If the
collision is assumed to be mediated by a long-lived complex
(due to the deep well in the PES), it is possible to
decompose the whole collision process in the step of
formation of the collision complex (capture), and the step
of its decomposition. Accordingly, the total (summed over
final states) reaction probability, $P_{r}^i(E)$  for
reagents colliding in the incoming channel $i$,
characterized by quantum numbers $(J,M,v,j,l)$, can be
broken down into two terms: i) the probability for the
colliding partners to be captured in the complex, $P_{\rm
capt}^i(E)$,  and ii) the probability for, once the complex
is formed, decomposing into the product arrangement
channel, $P_{\rm dec}^{ \to {\rm prod}}(E)$. In complex
mediated reactions with deep wells, the statistical {\em
ansatz} can be applied: if there existed a complete lost of
memory (randomization) of the reacting flux within the
complex, the break down of the complexes into fragments
would be independent of the initial state of the reagents
which originated them (except for the total angular
momentum and energy conservation). In such extreme case
(ergodic hypothesis), we can say that $P_{r}^i(E) \approx
P_{\rm capt}^i(E) \times P_{\rm dec}^{ \to {\rm prod}}(E)$.
Furthermore, the fraction of complexes which decompose into
the reactants (D$^{+}$+H$_2$) or products (H$^+$+HD) is
roughly proportional to the corresponding number of
energetically accessible channels available from the
complex, considering all of them as equiprobable.
Hereinafter, we will denote the fraction corresponding
to products, $P_{\rm dec}^{\to {\rm prod}}(E)$,
 as the {\em statistical factor}.  In
particular, if we denote with $A(E)$ and $B(E)$ the number
of energetically open channels corresponding, respectively,
to the reagent and product arrangements, the statistical factor
 may be approximated by
$B(E)/(A(E)+B(E))$ (number of favored outcomes divided by
the total number of equiprobable outcomes). For any  $J>1$
(see note~\footnote[4] {Let us consider, for example, the
case $J=2$. According to Fig.~\ref{fig0}, there are three
open rovibrational states of HD at the considered collision
energies. The incoming channel has quantum numbers ($v$=0,
$j$=0, $l$=2, $J$=2), and the total angular momentum,
$J=2$, and the parity, $\epsilon=(-1)^{j+l}=+1$ are
conserved in the collision; therefore, the product
collision channels which are coupled to the incoming
channel are ($v'$=0, $j'$=0, $l'$=2, $J$=2), ($v'$=0,
$j'$=1, $l'$=3, $J$=2), ($v'$=0, $j'$=1, $l'$=1, $J$=2),
($v'$=0, $j'$=2, $l'$=4, $J$=2), ($v'$=0, $j'$=2, $l'$=2,
$J$=2), ($v'$=0, $j'$=2, $l'$=0, $J$=2). This way, there
are 7 coupled channels, with $A(E)=1$ and $B(E)=6$.}) we
find that  $A(E)=1$ and $B(E)=6$, and one gets that
$B(E)/(A(E)+B(E))=6/7 \approx 86 \% $, the statistical
factor we have used to correct the Langevin result. However
for the particular cases $J=0$ and $J=1$ the fractions are
different, being 3/4 $\approx$ 75$\%$ and 5/6 $\approx$
83$\%$, as corresponds to 3 and 5 open product channels
(respectively) {\em versus} 1 open reactant channel. Let us
note that, if the number of product channels were very big,
$B(E)>>A(E)$, then $P_{\rm dec}^{ \to {\rm prod}}(E)\approx
1$ and the result would be equivalent to the Langevin
assumption. This quick reasonings can be put in solid
grounds by making use of more rigorous statistical
models\cite{Rack0}, in their quantum \cite{Rack1} or
quasiclassical versions \cite{Aoiztomas} applied in the
past to the H$^+_3$ system at thermal energies. They relate
the fraction of complexes which decompose into a particular
channel to the {\em capture} probability of forming the
complex starting from it. In our  naive reasoning, we are
assuming such capture probabilities to be $\approx 1$, what
has sense if we are not very close to threshold. We refer
to ref.~\cite{Lara2} for more details.

In Fig.~\ref{fig6}, the comparison of the obtained rate
coefficients with the experimental cross-sections
multiplied by the relative velocity from
Ref.~\cite{G:ACP92} is shown. The agreement is fairly good.
Note that the experimental results include also the
contribution of {\em ortho}-H$_2$. However, the latter
should not be very different to the one of {\em para}-H$_2$
according to recent calculations \cite{Honv2}.

Finally, it is interesting to note that the Langevin
expression $\sigma_{\rm L}(E)$, which should be valid only
for high energies (when many partial waves are open),
fulfills the Wigner threshold law ($\sigma_{\rm L} (E) \sim
E^{-1/2}$) in the limit of zero collision energy. Its slope
in Fig.~\ref{fig4} thus coincides with the slope
corresponding to our results. Such situation only takes
place for the  $n=4$ case (ion-induced dipole) and could be
considered a happy coincidence (for $n \ne $4, the
collision energy dependence of the cross-section is not any
longer $E^{-1/2}$). However, as can be seen in
Fig.~\ref{fig4}, also the absolute values of both curves
(the Langevin cross-sections and our calculated
cross-sections) agree remarkably well ($\sigma(E)\approx
1.07 \sigma_L (E)$) in the ultracold limit, where only one
partial wave is open. This fact deserves an explanation and
we  will devote the following subsection to rationalize it.


\subsection{
Low partial waves in the ultracold regime}

\begin{figure}
 \begin{center}
\includegraphics[width=50ex]{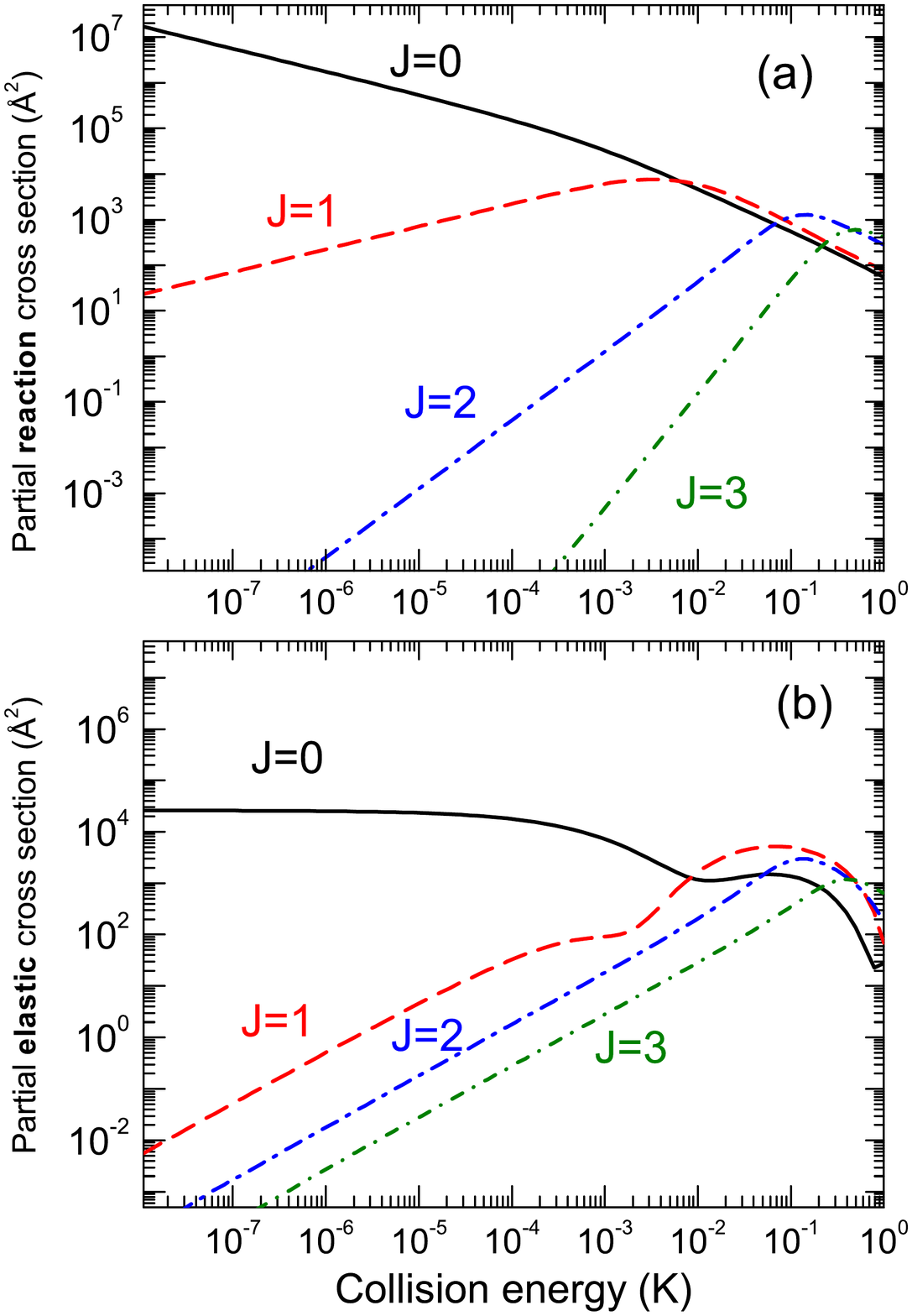}
\vspace{0.0cm}
\caption{Reaction and elastic cross-sections corresponding to
the lower partial waves for the collision D$^{+}$+H$_2$ ($v$=0, $j$=0) in the cold and ultracold regimes.
} \label{fig3}
\end{center}
\end{figure}

When analyzing the ultracold regime for a LR potential
behaving as  $-C_n/R^n$ $(n > 3)$, it is convenient to
define a characteristic length, $R_n=(2 \mu C_n
/\hbar^2)^{1/(n-2)}$, and a characteristic energy,
$E_n=\hbar^2 /(2 \mu R_n^{2})$. The latter is of the order
of the p-wave centrifugal barrier, which appears around
$R_n$. For the case $n$=4, $E_4$ actually coincides with
the height of the centrifugal barrier, assuming that the
latter is located far enough for the $-C_4/R^4$ asymptotic
behavior being valid. In our system, $R_4$ is approximately
99.7 a$_0$ and  $E_4 \approx 8.6\times10^{-3}$ K.

\begin{figure}
 \begin{center}
\includegraphics[width=55ex]{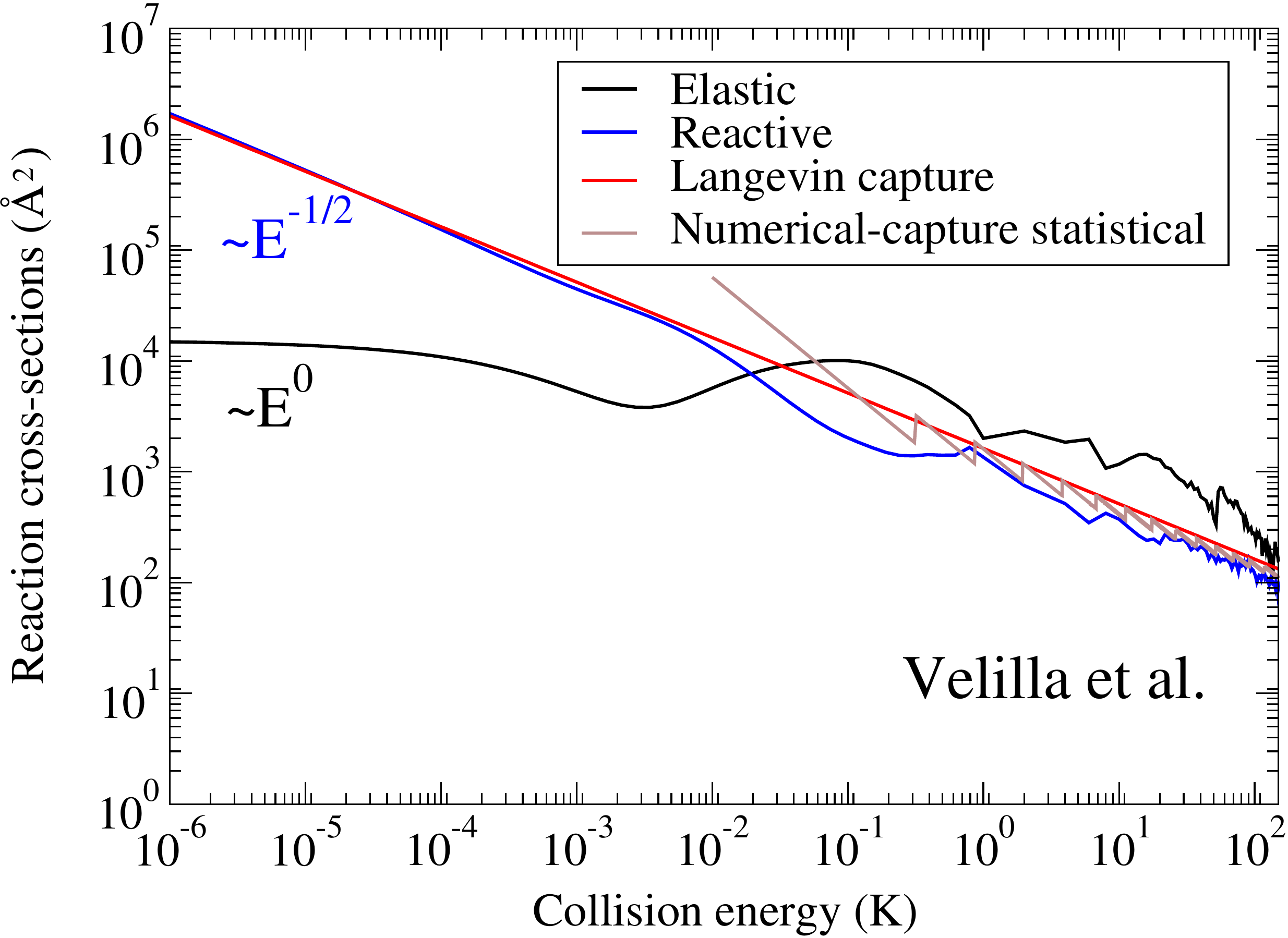}
\caption{Cross-sections for the collision D$^{+}$+H$_2$ ($v$=0, $j$=0).
Reaction cross-sections are compared with the Langevin prediction (for the right $C_4$)
and the results of a more realistic Numerical-capture statistical model.
} \label{fig4}
\end{center}
\end{figure}

\begin{figure}[t]
 \begin{center}
\includegraphics[width=55ex]{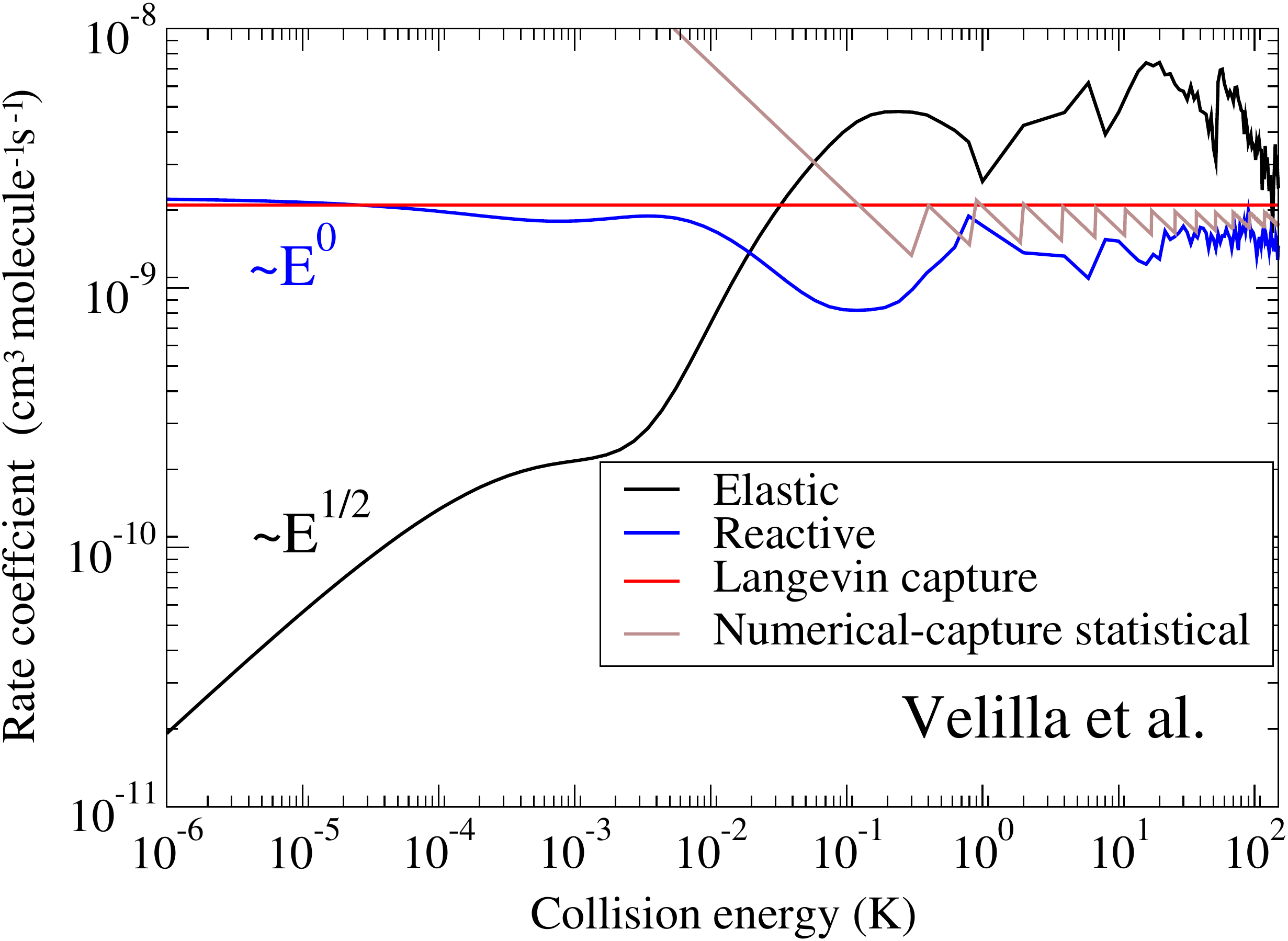}
\caption{Rate coefficients for the collision D$^{+}$+H$_2$ ($v$=0, $j$=0 on the PES by
Velilla {\em et al}. Reaction rates are compared with the Langevin prediction (for the right $C_4$)
and the results of a more realistic Numerical-capture statistical model.
} \label{fig5}
\end{center}
\end{figure}

\begin{figure}[t]
 \begin{center}
\includegraphics[width=50ex]{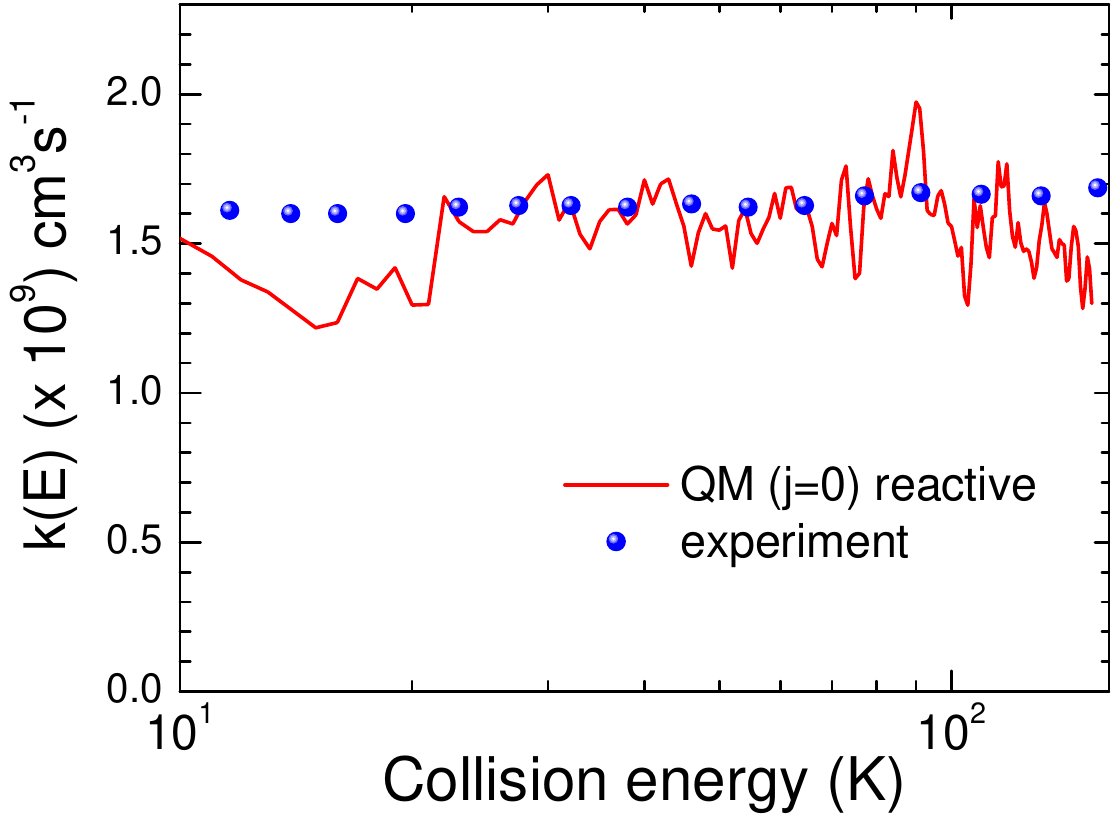}
\caption{Comparison of the rate coefficients for the collision D$^{+}$+H$_2$ ($v$=0, $j$=0) on the PES by Velilla {\em et al} with the experimental results from Ref.~\cite{G:ACP92}.
} \label{fig6}
\end{center}
\end{figure}

The calculations yield the S-matrix as a function of the
energy for each total angular momentum $J$ or,
equivalently, for each partial wave $l$ (for the case
$j=0$, both quantum numbers are equal and we will use them
interchangeably). This allows the calculation of an energy
dependent complex scattering length $\tilde{a}_{lm}(k)$ in
terms of the elastic (diagonal) element of the S-matrix,
\begin{equation}
\tilde{a}_{lm}(k)=\frac{1}{ik} \,\frac{1-S_{lm,lm}(k)}{1+S_{lm,lm}(k)}
\end{equation}
The definition of an energy dependent scattering length is
preferable to the use of effective range expansion, which
is not useful for very LR potentials like
$R^{-4}$~\cite{Simoni}. Let us denote with $\alpha_{lm}(k)$
and $\beta_{lm}(k)$ the real and imaginary part of the
complex scattering length:
$\tilde{a}_{lm}(k)=\alpha_{lm}(k)-i \beta_{lm}(k) $. At low
enough collision energy they relate to the elastic,
$\sigma_{\rm el}$, and total loss, $\sigma_{\rm loss}$
(inelastic plus reactive), cross-sections in the following
way:
\begin{eqnarray}\label{losse}
\sigma^{lm}_{\rm el} &\to& 4 \pi (\alpha_{lm}^2 + \beta_{lm}^2) \\
\sigma^{lm}_{\rm loss} &\to&  \frac{4\pi \beta_{lm}}{ k}
\end{eqnarray}
Given that H$_2$($v$=0, $j$=0) is the only rovibrational
state of the reactants open at the considered energies, the
inelastic process is absent and losses are only associated
to reaction, $\sigma_{\rm loss}^{l}=\sigma_{\rm re}^{l}$.
As we are not interested in product-state resolved
magnitudes, it is interesting to stress that all the
information that will be shown below is contained in the
diagonal elements $S_{lm,lm}(k)$:
\begin{eqnarray}
\sigma_{\rm el}&=&\sum_{l,m}\sigma^{lm}_{\rm el} = \frac{\pi}{k^2}\sum_{lm}|1-S_{lm,lm}(k)|^2 \\
\sigma_{\rm re}&=&\sum_{l,m}\sigma^{lm}_{\rm re} = \frac{\pi}{k^2}\sum_{lm}(1-|S_{lm,lm}(k)|^2)
\end{eqnarray}
The behavior of the cross-sections at very low kinetic
energies is given by the well known Wigner threshold
laws~\cite{Sade,Weiner}, which state that the elastic and
the total-loss cross-sections associated to each partial
wave vary close to threshold as:
\begin{eqnarray}
\sigma_{\rm el}^{l} &\sim&  E^{2l} \\
\sigma_{\rm loss}^{l} &\sim&  E^{l-1/2}
\end{eqnarray}
However, threshold laws for elastic scattering are modified
for a potential with $n=4$, having a pretty interesting
form~\cite{Sade,Weiner}. The behavior of the phase shift
for $ l > 0$ at very low collision energies is dominated by
a term $\sim E$  originating from the polarization
potential~\cite{Simoni}. This way, the behavior of the
elastic cross-section  is given by
\begin{eqnarray}
\sigma_{\rm el}^{0} &\sim&  \text{constant} \\
\sigma_{\rm el}^{l} &\sim&  E ~~\text{for}~ l>0
\end{eqnarray}
In summary, while the reaction cross-sections are expected
to change as $E^{l-1/2}$, the elastic cross-section will
remain constant for $l=0$, while changing as $\sim E$ for
any value of $l$ bigger than 0. These behaviors can be
distinguished in Fig.~\ref{fig3}, where the ultracold
reaction and elastic cross-sections for the lowest three
partial waves are shown. Note the opening of the p-wave
($J=l=1$) at energies around $E_4$. The considered
behaviors amount for a total reaction cross-section
changing as $E^{-1/2}$ (and reactive rate coefficient
constant) and total elastic cross-section reaching a
constant value (elastic rate coefficient changing as
$E^{1/2}$) in the limit of extremely low kinetic energies,
when only $l=0$ is open. Such behaviors can be clearly
distinguished in Figs.~\ref{fig4} and~\ref{fig5}.

As discussed above for cross-sections, and now switching to
rate coefficients, both the Langevin capture model and the
Wigner threshold laws predict constant reaction rates for
high and low kinetic energies (respectively) when applied
to the title system. The issue is how these constant values
relate. From our calculations, the limiting rate at low
energies almost coincides with the (classical) Langevin
value, the former being 1.07 times the latter (see
Fig.~\ref{fig5}). The result is a reaction rate which
remains basically constant for a range of more than eight
orders of magnitude! Variations would be further smoothed
by the Boltzmann averaging and, if confirmed by experiment,
an essentially constant thermal rate would result.  However
the ``Langevin behavior'' is expected when many partial
waves are open. What is then  the reason for this
``Langevin behavior'' when there is only one? What is the
reason for this 'universality', understood in the sense of
an exclusive dependence on the LR interactions (the only
ones considered in the capture model)? However paradoxical
the situation may appear, collisions at very low kinetic
energy favor the LR interactions. The question should
probably be the reverse: how is it possible that a capture
model, based on LR interactions, is commonly considered in
reaction dynamics at high energies?
Part of the answer is the following: once we assume that all the flux that
reaches the SR region of the PES leads to reaction, we implicitly
eliminate any effect that the details of SR region may have on the outcome,
and LR is the only thing that remains.

Very recently, {\em quantal versions} of the Langevin model
have been proposed in two different
approaches~\cite{Jach:03,Gao1}.
Starting from slightly different quantum defect theory
frameworks, universal models are obtained which change
smoothly from the ultracold to the high energy regime under
the same Langevin assumption:
all the flux that reaches the short-range (whose amount is
controlled by the LR interactions) leads to reaction; or,
in other words, the loss probability at SR is the unity.
The conclusion derived from these models for the case $n$=4
is the following: in the limit of zero kinetic energy, the
reactive cross-section should reach a value $2 \sigma_{\rm
L}(E)$, that is, twice that predicted by the Langevin
expression and not simply $\sigma_{\rm L}(E)$, which is
what we essentially get. As stressed by the authors, such
model represents a type of universal behavior which can
emerge whenever the number of open channels in a set of
coupled channel equations becomes large. However, as
already noted, this is not the case for the title reaction,
where only 4 channels (three product channels and one
reactant channel) are open for $J$=0. Luckily, the
formalism in Ref.~\cite{Jach:02} generalizes the model in
Ref.~\cite{Jach:03} to the non-universal regime where the
loss probability at SR is less than the unity ($P^{\rm re}<
1$). This model will be applied to our reaction in the
following section.

\subsection{The scattering lengths}

In this section, we will apply the quantum model for
ultracold reactive collisions in $1/R^n$ potentials,
published in ref.~\cite{Jach:03}, to rationalize the
results we have obtained in the ultracold regime. The model
assumes the formulation of multichannel quantum defect
theory (MQDT) by Mies~\cite{Mies1,Mies2} to find a general
expression for the complex scattering length
$\tilde{a}_{lm}(k)$ as a function of the MQDT functions.
This allows to parametrize the dependence of the complex
scattering length with $k$ in terms of two real parameters,
$y$ and $s$, together with the mean scattering length,
$\bar{a}$~\cite{Griba}. The latter is defined as
\begin{eqnarray}\label{meanscat}
\bar{a} = \frac{\pi (n-2)^{(n-4)/(n-2)}}{\Gamma ^2 (1/(n-2))}R_n
\end{eqnarray}
where $R_n$ was defined above. In the case $(n=4)$
$\bar{a}$ coincides with  $R_4$ and is thus given by
$\bar{a}= (2 \mu C_4 )^{1/2}/ \hbar$ ($\approx$99.7 a$_0$).

The dimensionless parameter $0 \le y \le 1$, the SR
inter-channel coupling strength, characterizes the SR
collision. It relates to $P^{\rm re}$, the probability of
irreversible loss of flux from the entrance channel due to
the dynamics at SR, according to $P^{\rm re}=4y/(1+y)^2$.
The case $y=1$ corresponds to the Langevin assumption or
'universal' case, where all the flux that reaches the SR
region leads to reaction.

The dimensionless scattering length $s=a/ \bar{a}$ is an
entrance channel phase, where $a$ is the ``background''
s-wave scattering length corresponding to the one-channel
potential.  It is related to the phase of the standing wave
due to the interference between the incoming and outgoing
waves in the outer region~\cite{Jach:02,Jach:03}. The
limits $s \rightarrow \pm \infty$ correspond to a bound
state crossing threshold.

The model has been checked already for a few reactive
systems~\cite{Jach:02,Jach:03}. In particular, it has been
found able to satisfactorily fit the measured reaction
rates for the Penning ionization of Ar by metastable He in
a wide range of collision energies over
threshold~\cite{Jach:02}, by simply assuming the same
values of $s$ and $y$ for all partial waves.

In terms of the parameters of the model, the limit values
for the real and imaginary parts of the s-wave ($l$=0)
complex scattering length are given
by~\footnote[5]{Equations (\ref{l0}), (\ref{l0b}),
(\ref{l1}), (\ref{l1b}) were kindly provided by the authors
of ref.~\cite{Jach:02}; Equations (\ref{l2}), (\ref{l2b}),
(\ref{l3}), (\ref{l3b}), were deduced by the authors of
this work following the formalism in ref.~\cite{Jach:02},
and according to the information in ref.~\cite{Simoni}.}):
\begin{eqnarray}\label{l0}
\alpha_{00}(k) &\to&  \bar a \frac{s(1-y^2)}{1+s^2y^2} \\ \label{l0b}
\beta_{00}(k) &\to& \bar a \frac{y(1+s^2)}{1+s^2y^2}
\end{eqnarray}
Regarding $l=1$, the corresponding predictions are as follows:
\begin{eqnarray}\label{l1}
\alpha_{1m}(k)  &\to&  -k \bar{a}^2 \pi /15  \\ \label{l1b}
\beta_{1m}(k) &\to&  \bar{V} k^2 \frac{y(1+s^2)}{s^2+y^2}
\end{eqnarray}
where the characteristic {\em p}-wave volume  for $n$=4 is
defined as $\bar{V}=\bar{a}^3/9$.

\begin{figure}[t]
 \begin{center}
\includegraphics[width=65ex]{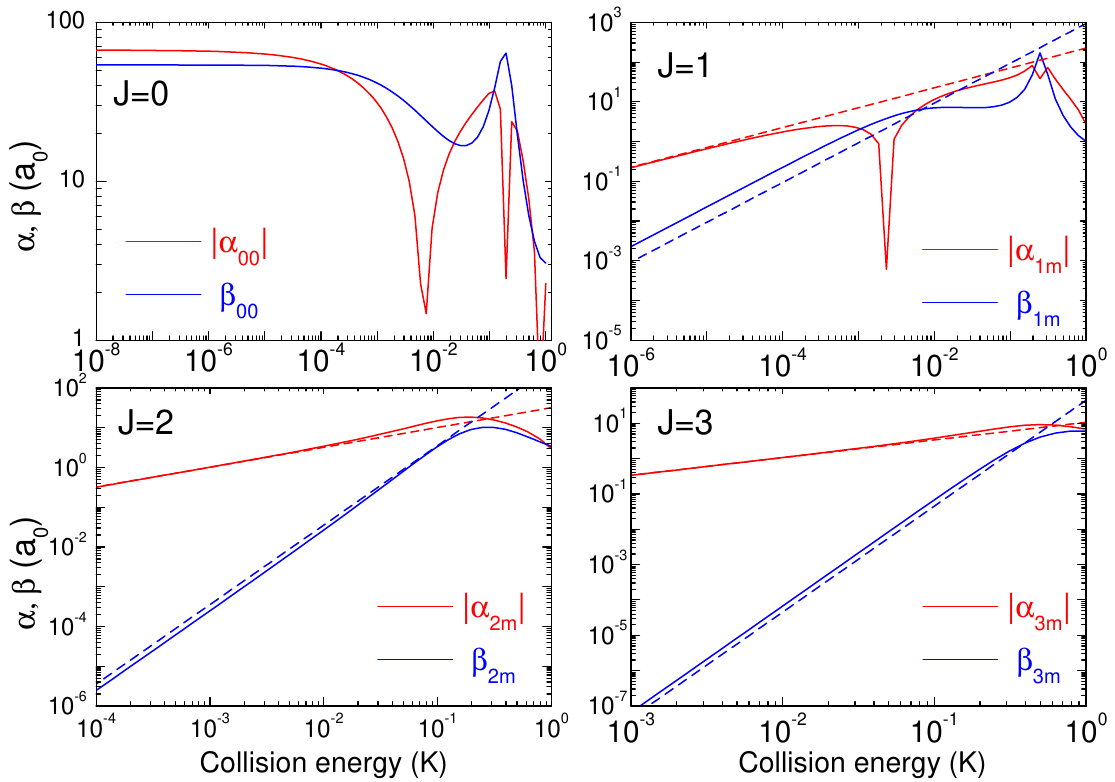}
\caption{Real and imaginary parts of the scattering lengths obtained in the
calculations for the 4 lowest partial waves. The scale is logarithmic and the absolute
value of $\alpha$, which is mostly negative, is plotted. The values for $J=$1, 2 and 3
(in continuous line) are compared with the predictions from the model in Ref.~\cite{Jach:02}
(in dashed lines), calculated using
Eqs.~\eqref{l1}-\eqref{l3b}, assuming  for $s$ and $y$ the same values that
have been obtained for $J=0$.
} \label{fig7}
\end{center}
\end{figure}

With regard to $l$=2,
\begin{eqnarray}\label{l2}
\alpha_{2m}(k)  &\to&  -k \bar{a}^2 \pi /105  \\ \label{l2b}
\beta_{2m}(k)  &\to&  k^4 \bar{a}^5 \frac{y(1+s^2)}{2025(1+s^2y^2)}
\end{eqnarray}
Finally, for $l$=3,
\begin{eqnarray}\label{l3}
\alpha_{3m}(k)  &\to&  -k \bar{a}^2 \pi /315  \\ \label{l3b}
\beta_{3m}(k)  &\to&  k^6 \bar{a}^7 \frac{y(1+s^2)}{2480625(s^2+y^2)}
\end{eqnarray}
We have analyzed the scattering lengths obtained from the
calculations in  terms of these expressions. The upper-left
panel of Fig.~\ref{fig7} depicts the real and imaginary
parts of the scattering length corresponding to $J=0$
(obtained using the element $S_{lm,lm}=S_{00,00}$ from our
calculations) as a function of the collision energy. Both
reach constant values in the ultracold regime, in agreement
with the Wigner threshold laws discussed in the previous
subsection and with the fact that Eq.~\eqref{l0} and
Eq.~\eqref{l0} are independent on $k$.  The real and
imaginary parts of the scattering lengths corresponding to
the three following partial waves, $J=1$, $J=2$ and $J=3$
are also shown in the other three panels of the
Fig.~\ref{fig7}. The limiting behaviors as a function of
the energy are again in perfect agreement with the expected
threshold laws and the power of the dependence on $k$ of
the expressions \eqref{l0}-\eqref{l3b}.

In order to parametrize our results in terms of $s$ and
$y$, let us consider first the case $l=0$ ($J=0$).
Introducing the limiting values we have obtained in the lhs
of Eq.~\eqref{l0} and Eq.~\eqref{l0b} and solving for the
parameters $y$ and $s$, we obtain $y(l=0) = 0.34$ ($P^{\rm
re} = 77 \% \approx 3/4$) and $s(l=0) = -0.8$. The
parametrization for higher values of $l$ ($J$) is not so
simple and we will consider real and imaginary parts
separately. Wigner threshold laws are already modified
starting from $l=1$. The term that dominates the behavior
of $\alpha_{lm}(k)$ for $l>0$ does not depend on $y$ or
$s$, and the real parts, given by Eqs.~\eqref{l1},
\eqref{l2} and \eqref{l3}, can be compared directly with
our scattering results. The values for $\alpha$ in the
three considered cases agree very well (within $1\%$) with
those given by the expressions \eqref{l1}, \eqref{l2}
\eqref{l3}. This coincidence can be considered a test of
the theory and serves to ensure the convergence of the
(computationally demanding) scattering calculations. It
should be remarked that these expressions do not depend on
$s$ or $y$, and hence we can consider them as really
universal, being only dependent on the $C_4$ and thus of
the LR part of the potential exclusively.

The analysis of the imaginary parts given by
Eqns.~\eqref{l1b}, \eqref{l2b}, and \eqref{l3b} is not so
straightforward. Given that the real part of the scattering
length, $\alpha_{lm}(k)$, does not depend on $s(l)$ and
$y(l)$, it is not possible to solve for the two unknown
quantities only using the expression of $\beta_{lm}(k)$.
Similarly to the procedure followed in
ref.~\onlinecite{Jach:02}, what can be done instead is to
assume that $y$ and $s$ do not strongly depend on $l$ and
check if the values for $\beta_{l,m}$, obtained by
substituting $y(l=0)$ and $s(l=0)$ in Eqs.~\eqref{l1b},
\eqref{l2b} and \eqref{l3b} are not very different from
those obtained in the scattering calculations. After doing
that, we find that the ratios of the $\beta_{l,m}$'s we
obtain substituting $y(l=0)$ and $s(l=0)$ in the
expressions of the model, and the accurate ones, extracted
from our scattering calculations, are: $0.4$ for $J=1$,
$1.4$ for $J=2$ and $0.7$ for $J=3$. Therefore, the
agreement seems to be acceptable on average. This is likely
one of the reasons that the assumption of the same values
of $s$ and $y$ for different partial waves was found to
work well for the Penning ionization of Ar by metastable
He. In such a system the reduced mass is bigger (four
times) than in present case, what increases the number of
partial waves considered, and hence probably reducing the
average error of such strong approximation. A necessary
condition (but not sufficient) for the same values of
$s(l=0)$ and $y(l=0)$ to be valid for other partial waves
is that the $(n=4)$ behavior  in the potential extends from
the asymptote down to the region where the corresponding
centrifugal barriers are located. This condition is
fulfilled in this case for the considered partial waves
whose centrifugal barriers lie around 99 a$_0$, 57 a$_0$,
and 41 a$_0$, respectively. In this way, the expressions of
the model could, in principle,  work well with values for
the parameters that do not strongly depend on $l$. However,
it is interesting to note that the small number of states
which are coupled to the incident channel changes when
moving from $J=0$ to higher values of $J$, as explained
above. This way, differences in the parameters, in
particular in $P^{\rm re}$ and thus in $y$, are not
unexpected.

To the best of our knowledge, the values for y ($P^{\rm
re}$) and $s$ in the considered model are considered
phenomenological. To date they have been obtained for very
particular systems by fitting experimental results to the
analytical expressions of the model~\cite{Jach:02}
(equivalent to what we have done with our theoretical
results to extract $s$ and $y$). However, we have some
clues to rationalize the value for $y(l=0)$ that we have
finally found; in fact, our previous expectations where
confirmed with such value.
The so-called universal case ($y=1$), usually considered as
a first approach to any system in recent literature, may be
characterized for a number of coupled states which is so
huge that one may assume that all the reactive flux is
irreversibly lost from the incident channel. However, even
if the number of coupled channels inside the well is very
large for the title system, its exothermicity is low (the
difference in zero point energies) and consequently, the
number of open rovibrational states is small (three for the
HD product and $J$=0) and hence the number of open channels
coupled to the incident is also small. This anticipates the
possibility, discussed above, of a significant flux
returning from the complex (recrossing) to the incident
channel. According to this, it would be possible to relate
the value obtained for $P^{\rm re}$ with the statistical
factor 3/4 corresponding to $J=0$. If the randomization and
lost of memory of the system at SR (in the complex) were
complete, a $P^{\rm re}=3/4$, as obtained, would be
reasonable. A value of $s=-0.8$, very close to -1, would
guarantee that resonant states close to threshold were not
masking the equiprobable distribution among states that, on
average,  the presumably long residence time within the
well would imply.

However, the situation is not so simple. The very good
agreement between the statistical factor and $P^{\rm re}$
might simply be fortuitous. Actually, it is difficult to
derive an unambiguous proportionality between $P^{\rm re}$
and the statistical factor. Calculations with $J=0$
 where the mass of
the deuterium is artificially modified in order to change
the number of open rovibrational states of HD
(thus modifying $B(E)/(A(E)+B(E))$),
give rise to a value of $P^{\rm re}$ which keeps some
correlation with the corresponding statistical factor.
However this value can also widely oscillate, and
the oscillations are usually found in coincidence with bigger values
of $s$, which can be associated to resonances. Actually, as we
will discuss hereunder, it is expected that the local
effect of resonances  can modify  the average behavior
expressed by the statistical factor.


More work has to be performed to conclude the average
agreement between the values of $P^{\rm re}$ and
statistical factors in systems with a well. However, let us
note that we are considering a real system which only
approximately fulfills all the assumptions in the model. We
are using expressions \eqref{l0} and \eqref{l0b},
presumably only valid for an ideal case, to extract
particular values of parameters whose meaning is only
defined in that ideal case. It is difficult to predict how
this divergence from ideality is going to affect not only
the value for $s$ and $y$ but also their meaning. The pure
roles of $y$ and $s$ are probably difficult to unravel for
real systems.

For the title reaction, statistical models seem to
yield results in good agreement with accurate, fully
converged quantum scattering for thermal reactions rates
even at low temperatures~\cite{Honv1}. As an instance, in
Fig.~1 of ref.~\citenum{Honv1} it is shown that
the statistical model by Rackham and Manolopoulos \cite{Rack0}
leads to a step-like reaction probability that
on average reproduces the highly structured one found in
the quantum calculations. Since for thermal rate
coefficients only the average value over a range of
energies matter, the statistical model has proved to render
accurate rate coefficient values. However, it remains to be
seen if these models can work in the ultracold regime where
a coarse grid hypothesis is not longer valid. Moreover,  by
its nature, the statistical models would fail to describe
the detailed effect of resonances, which are ubiquitous as
long as it exists a sufficiently deep well. Given the
resonant nature of the behavior close to threshold, the
statistical model may find unsurmountable obstacles in the
cold and ultracold regimes. Implicit in the statistical
model is the assumption of a lost of coherence between
external and internal regions of the potential; the
presence of a resonance, which means an increase of
amplitude in the inner region, would invalidate such an
assumption preventing the factorization $P_{r}^i(E) \approx
P_{\rm capt}^i (E) \times P_{\rm dec}^{ \to {\rm prod}}(E)$
that was discussed above.

In this regard, the structure of the previous expression is
similar to the {\em ansatz} $P_{\rm r}^i(E) \approx  P_{\rm
LR} (E) \times P_{\rm SR}(E)$, which is at the conceptual
basis of some approximate approaches to ultracold reactions
recently proposed~\cite{Fara}. This {\em ansatz} is inspired by the
MQDT concept of factoring the S-matrix into separate parts
due to the short and LR interactions.
Accordingly, the flux which reacts is only the fraction
$P_{\rm SR}(E)$ (determined by the couplings at SR) of the
total transmitted by the LR interactions, $P_{\rm LR} (E)$.
This seemingly naive approach can be put into question by
the non-local nature of the wave function, especially when
resonant states lie close to threshold. This can be
concluded from the results of some other models
\cite{Gao2008,Koto2012,Jach:02} and, in particular, by
using Eq.~\eqref{l0b}. From this equation and applying
expression \eqref{losse}, it can be seen that the reaction
probability at ultracold energies is given by
\begin{equation}
P_{\rm r}^i(E)=4 k\beta_{00}(k)= 4k\bar{a} \times
\frac{y(1+s^2)}{1+s^2y^2}
\end{equation}
The first factor can be recognized as the capture, or
probability transmitted by the LR potential (and thus
identified with $P_{\rm LR} (E)$~\footnote[6]{This capture
probabilities can be numerically calculated for any value
of $k$ by propagating inwards the logarithmic derivatives
of regular and irregular Bessel functions under the effect
of the LR potential and matching at short distance with
perfect absorption WKB conditions. The value of these
capture probabilities at very low $k$ is given by
$4k\bar{a}$}), but the second, the one we would associate
with the statistical factor, does not coincide with the
$P^{\rm re}$ defined by the model. Moreover, the second
term can be paradoxically bigger that 1 for low values of
$y$ and big values of $s$ (that is, for resonant states
close to threshold), so it cannot be associated in general
to a probability. For the particular case of unit loss
probability at short range, one gets $P_{\rm
r}^i(E)=4k\bar{a} \times 1$, what is consistent with the
{\em ansatz} and with the intuitive, but wrong, conclusion
that the reaction probability is limited by the amount of
probability that the LR is able to capture. Also for the
case $s=1$ (which can be associated with minimum influence
of resonant states), the expression given by the model
\begin{equation}
P_{\rm r}^i(E)=4k\bar{a} \times \frac{2y}{1+y^2}
\end{equation}
is approximately equal to
\begin{equation}
P_{\rm r}^i(E)=4k\bar{a} \times \frac{4y}{(1+y)^2}
\end{equation}
but only for $y \approx 1$~\footnote[6]{This can be seen
assuming $y=1-x$ and expanding in powers of $x$ for
$x<<1$}; for a general value of $y$, differences between
both expressions can be as big as $25 \%$. However, for
general large values of $s$ and small values of $y$  the
factor $y(1+s^2)/(1+s^2y^2)$ can be bigger than one,
leading to reaction probabilities which are bigger than the
capture; it is thus difficult to conceive how a statistical
factor could account for it. We have checked this extreme case again by
artificially changing the mass of the deuterium. For some
particular values of the mass, the reaction probability is
found much bigger than the capture  by the potential in the
reactant channel, $4k\bar{a}$. This is found, for example,
for a value of $s=-3.98$.  Let us finally note that
the statistical model, in the way that it is usually implemented,
calculates the capture probability of a particular state
considering its coupling with other states in the entrance
channel \cite{Rack0}, thus providing a {\em dynamical} value
for the capture which may be
different than the one-channel capture $P_{\rm
r}^i(E)=4k\bar{a}$.  However, this reasoning calls our attention
to possible limits for applicability of the statistical
model to very low temperatures or kinetic energies in the
presence of resonant states close to threshold.

\section{Summary and Conclusions}
We have carried out a detailed study of the D$^+$+H$_2$
ion-molecule system under the cold and ultracold regimes,
covering the 10$^{-8}$--150 K collision energy range.
Rigorous fully converged quantum mechanical calculations
have been performed on the potential energy surface by
Velilla {\em et. al}, which faithfully reproduces the long
range (quadrupole and charge-induce dipole) behavior. The
quantum dynamical methodology used in this work is
especially appropriate to calculate {\em ab initio} cross
sections at extremely low collision energies. In addition,
the method is ideally suited to tackle complex mediated
reactions.

The methodology has been applied for the first time to a
reactive ion-molecule system governed by a
$R^{-4}$ potential at long range. This makes the calculations more
demanding since propagations up to distances on the order
of 10$^5$ a$_0$ are required.  The calculated elastic and
reactive cross sections and rate coefficients have been found to
comply with the
expected threshold laws. Interestingly, the behavior in the
ultracold regime is well described by the classical
Langevin model, which is expect to work only when many
partial waves are open. Should this behavior be
confirmed by the experiment, it would have been found a
system with a reaction rate coefficient which remains
almost constant in a kinetic energy range of more than ten orders of
magnitude.

The detailed calculated reaction probabilities
 and the associated scattering lengths have allowed us
to test the quantum defect theory model by Jachymsky {\em
et al.} for various partial
waves. The assumption made in previous works that model
parameters are independent on the specific partial wave has
been assessed with the present calculations. The scattering
lengths calculated under this assumption are found in
acceptable, at least on average, agreement with the
scattering results.

In addition, rough statistical assumptions have been found
able to rationalize the obtained loss probability at
short range. Although there seems to be a correlation
between the statistical factor (corresponding to the statistical
model) and the loss
probability at short range (corresponding to the
model by Jachymsky {\em et al.}) in the case of this process,
mediated by a deep well, more work is needed to fully
evaluate it.

\begin{acknowledgments}
The authors are greatly indebted to K. Jachymski for
fruitful discussions and for providing us with the
expressions of the complex scattering length for $l$=0 and
$l$=1 partial waves. The Spanish Ministries of Science and
Innovation and Economy and Competitiveness (grants
CTQ2008-02578/BQU, CSD2009-00038, and CTQ2012-37404-C02)
are gratefully acknowledged.
\end{acknowledgments}




\begin{thebibliography}{99}
\providecommand{\natexlab}[1]{#1}
\providecommand{\url}[1]{\texttt{#1}}
\expandafter\ifx\csname urlstyle\endcsname\relax
  \providecommand{\doi}[1]{doi: #1}\else
  \providecommand{\doi}{doi: \begingroup \urlstyle{rm}\Url}\fi

\bibitem[Aguado and Paniagua(1992)]{Agua} A.~Aguado and
    M.~Paniagua.
\newblock \emph{J. Chem. Phys.}, 96:\penalty0 1265, 1992.

\bibitem[Aguado et~al.({2000})Aguado, Roncero, Tablero,
    Sanz, and
  Paniagua]{ARTSP:JCP00}
A.~Aguado, O.~Roncero, C.~Tablero, C.~Sanz, and
M.~Paniagua.
\newblock \emph{{J. Chem. Phys.}}, {112}:\penalty0 {1240}, {2000}.

\bibitem[Aoiz et~al.(2008)Aoiz, Gonz\'{a}lez-Lezana, and
  R\'{a}banos]{Aoiztomas}
F.~Aoiz, T.~Gonz\'{a}lez-Lezana, and V.~S. R\'{a}banos.
\newblock \emph{J. Chem. Phys.}, 129:\penalty0 094305, 2008.

\bibitem[Aoiz et~al.(2007{\natexlab{a}})Aoiz,
    {Gonz\'alez-Lezana}, and
  {S\'aez--R\'abanos}]{AGS:JCP07}
F.~J. Aoiz, T.~{Gonz\'alez-Lezana}, and
V.~{S\'aez--R\'abanos}.
\newblock \emph{{J. Chem. Phys.}}, {127}:\penalty0 {174109},
  2007{\natexlab{a}}.
\newblock \doi{{10.1063/1.2774982}}.

\bibitem[Aoiz et~al.(2007{\natexlab{b}})Aoiz, {S\'aez
    R\'abanos},
  Gonz\'alez-Lezana, and Manolopoulos]{ASGM:JCP07}
F.~J. Aoiz, V.~{S\'aez R\'abanos}, T.~Gonz\'alez-Lezana,
and D.~E.
  Manolopoulos.
\newblock \emph{{J. Chem. Phys.}}, 126:\penalty0 161101, 2007{\natexlab{b}}.
\newblock \doi{{10.1063/1.2723067}}.

\bibitem[Bachorz et~al.(2009)Bachorz, Cenek, Jaquet, and
    Komasa]{BCJK:JCP09} R.~A. Bachorz, W.~Cenek, R.~Jaquet,
    and J.~Komasa.
\newblock {}.
\newblock \emph{{J. Chem. Phys.}}, 131:\penalty0 024105, 2009.

\bibitem[Berblinger and Schlier(1994)]{BS:JCP94}
    M.~Berblinger and C.~Schlier.
\newblock \emph{{J. Chem. Phys.}}, 101:\penalty0 4750, 1994.

\bibitem[Berteloite et~al.(2009)Berteloite, Lara, Picard,
    Dayou, Launay,
  Canosa, and Sims]{nuestroFara}
C.~Berteloite, M.~Lara, S.~D.~L. Picard, F.~Dayou, J.-M.
Launay, A.~Canosa, and
  I.~R. Sims.
\newblock \emph{Faraday Discuss.}, 142:\penalty0 236, General dicussion, 2009.

\bibitem[Canosa et~al.(2008)Canosa, Goulay, Sims, and
    Rowe]{Canosa} A.~Canosa, F.~Goulay, I.~R. Sims, and
    B.~R. Rowe.
\newblock \emph{Low Temperatures and Cold Molecules}.
\newblock World Scientific, Singapore, 2008.

\bibitem[Carmona-Novillo et~al.({2008})Carmona-Novillo,
    {Gonz\'alez-Lezana},
  Roncero, Honvault, Launay, Bulut, Aoiz, {Ba\~nares}, Trottier, and
  Wrede]{CGRHLBABTW:JCP08}
E.~Carmona-Novillo, T.~{Gonz\'alez-Lezana}, O.~Roncero,
P.~Honvault, J.-M.
  Launay, N.~Bulut, F.~J. Aoiz, L.~{Ba\~nares}, A.~Trottier, and E.~Wrede.
\newblock \emph{{J. Chem. Phys.}}, {128}:\penalty0 014304, {2008}.

\bibitem[Costes and Naulin(2010)]{Costes2} M.~Costes and
    C.~Naulin.
\newblock \emph{Phys. Chem. Chem. Phys}, 12:\penalty0 9154, 2010.

\bibitem[Dulitz et~al.(2014)Dulitz, Motsch, Vanhaecke, and
    Softley]{Zeem} K.~Dulitz, M.~Motsch, N.~Vanhaecke, and
    T.~P. Softley.
\newblock \emph{J. Chem. Phys.}, 140:\penalty0 104201, 2014.

\bibitem[Fehsenfeld et~al.(1974)Fehsenfeld, Albritton,
    Bush, Fournier, Govers,
  and Fournier]{FABFGF:JCP74}
F.~C. Fehsenfeld, D.~L. Albritton, Y.~A. Bush, P.~G.
Fournier, T.~R. Govers,
  and J.~Fournier.
\newblock \emph{{J. Chem. Phys.}}, 61:\penalty0 2150, 1974.

\bibitem[Gao(2008)]{Gao2008} B.~Gao.
\newblock \emph{Phys. Rev. A}, 78:\penalty0 012702, 2008.

\bibitem[Gao(2011)]{Gao1} B.~Gao.
\newblock \emph{Phys. Rev. A}, 83:\penalty0 062712, 2011.

\bibitem[Geppert et~al.(2004)Geppert, Goulay, Naulin,
    Costes, Canosa, Picard,
  and Rowe]{Costes}
W.~D. Geppert, F.~Goulay, C.~Naulin, M.~Costes, A.~Canosa,
S.~D.~L. Picard, and
  B.~R. Rowe.
\newblock \emph{Phys. Chem. Chem. Phys}, 6:\penalty0 566, 2004.

\bibitem[Gerlich(1982)]{G:SASP82} D.~Gerlich.
\newblock \emph{{\rm in} {Symposium on Atomic and surface Physics.} {\rm Pg.
  304}, {\rm Eds. W. Lindinger, F. Howorka, T. D. M\"ark, and F. Egger}}.
\newblock Institut fuer Atomphysik der Universitat Innsbruck, Innsbruck, 1982.

\bibitem[Gerlich(1992)]{G:ACP92} D.~Gerlich.
\newblock \emph{{Adv. Chem. Phys.}}, 82:\penalty0 1, 1992.

\bibitem[Gerlich(1993)]{G:JCSFAR93} D.~Gerlich.
\newblock \emph{{J. Chem. Soc. Faraday Trans.}}, 89:\penalty0 2199, 1993.

\bibitem[Gerlich(1995)]{G:PS95} D.~Gerlich.
\newblock \emph{{Phys. Scripta}}, T59:\penalty0 256, 1995.

\bibitem[Gerlich and Schlemmer(2002)]{GS:PSS02} D.~Gerlich
    and S.~Schlemmer.
\newblock \emph{{Planet. Space Sci.}}, 50:\penalty0 1287, 2002.

\bibitem[Gerlich et~al.(1980)Gerlich, Nowotny, Schlier, and
    Teloy]{GNST:CP80} D.~Gerlich, U.~Nowotny, C.~Schlier,
    and E.~Teloy.
\newblock Complex-formation in proton-d2 collisions.
\newblock \emph{{Chem. Phys.}}, 47:\penalty0 245, 1980.

\bibitem[Gerlich et~al.(2013)Gerlich, sil, Zymak, Hejduk,
    Jusko, Mulin, and
  Glos\'ik]{Gerl}
D.~Gerlich, R.~P. sil, I.~Zymak, M.~Hejduk, P.~Jusko,
D.~Mulin, and
  J.~Glos\'ik.
\newblock \emph{J. Phys. Chem. A}, 117:\penalty0 10068, 2013.

\bibitem[{Gonz\'alez-Lenzana}
    et~al.(2009){Gonz\'alez-Lenzana}, Honvault,
  Jambrina, Aoiz, and Launay]{GHJAL:JCP09}
T.~{Gonz\'alez-Lenzana}, P.~Honvault, P.~G. Jambrina, F.~J.
Aoiz, and J.-M.
  Launay.
\newblock \emph{{J. Chem. Phys.}}, 131:\penalty0 044315, 2009.

\bibitem[{Gonz\'alez-Lezana}
    et~al.(2005){Gonz\'alez-Lezana}, Aguado, Paniagua,
  and Roncero]{GAPR:JCP05}
T.~{Gonz\'alez-Lezana}, A.~Aguado, M.~Paniagua, and
O.~Roncero.
\newblock \emph{{J. Chem. Phys.}}, 123:\penalty0 194309, 2005.

\bibitem[Gonz\'alez-Lezana et~al.(2013)Gonz\'alez-Lezana,
    Honvault, and
  Scribano]{Honv1}
T.~Gonz\'alez-Lezana, P.~Honvault, and Y.~Scribano.
\newblock \emph{J. Chem. Phys.}, 139:\penalty0 054301, 2013.

\bibitem[Gribakin and Flambaum(1993)]{Griba} G.~F. Gribakin
    and V.~V. Flambaum.
\newblock \emph{Phys. Rev. A}, 48:\penalty0 546, 1993.

\bibitem[Grozdanov and McCarroll(2011)]{GM:JPCA11} T.~P.
    Grozdanov and R.~McCarroll.
\newblock \emph{{J. Phys. Chem. A}}, 115:\penalty0 6872, 2011.
\newblock \doi{10.1021/jp1115228}.

\bibitem[Hawley et~al.({1990})Hawley, Mazely, Randeniya,
    Smith, Zeng, and
  Smith]{HMRSZS:IJMSIP90}
M.~Hawley, T.~L. Mazely, L.~K. Randeniya, R.~S. Smith,
X.~K. Zeng, and M.~A.
  Smith.
\newblock \emph{{Int. J. Mass Spectrom. Ion. Process.}}, {97}:\penalty0 {55},
  {1990}.

\bibitem[Henchman et~al.(1981)Henchman, Adams, and
    Smith]{HAS:JCP81} M.~J. Henchman, N.~G. Adams, and
    D.~Smith.
\newblock \emph{{J. Chem. Phys.}}, 75:\penalty0 1201, 1981.

\bibitem[Henson et~al.(2012)Henson, Gersten, Shagam,
    Narevicius, and
  Narevicius]{Nar1}
A.~B. Henson, S.~Gersten, Y.~Shagam, J.~Narevicius, and
E.~Narevicius.
\newblock \emph{Science}, 338:\penalty0 234, 2012.

\bibitem[Herbst(2001)]{H:CSR01} E.~Herbst.
\newblock \emph{{Chem. Soc. Rev.}}, 30:\penalty0 168, 2001.

\bibitem[Hollmann and Pigarov(2002)]{HP:PP02} E.~M.
    Hollmann and A.~Y. Pigarov.
\newblock \emph{{Phys. Plasmas}}, 9:\penalty0 4330, 2002.

\bibitem[Honvault and Dynamics(2004)]{hon04} P.~Honvault
    and J.-M.~L. Dynamics.
\newblock \emph{in {\sl Theory of Chemical Reaction Dynamics}}.
\newblock NATO Science Series vol. 145, Kluwer, 2004.

\bibitem[Honvault and Scribano(2013{\natexlab{a}})]{Honv2}
    P.~Honvault and Y.~Scribano.
\newblock \emph{J. Phys. Chem. A}, 117:\penalty0 9778, 2013{\natexlab{a}}.

\bibitem[Honvault and
    Scribano(2013{\natexlab{b}})]{Honverr} P.~Honvault and
    Y.~Scribano.
\newblock \emph{J. Phys. Chem. A}, 117:\penalty0 13205, 2013{\natexlab{b}}.

\bibitem[Hudson et~al.(2008)Hudson, Gilfoy, Kotochigova,
    and andD.
  DeMille]{hudsonexp}
E.~R. Hudson, N.~B. Gilfoy, S.~Kotochigova, and J.~M.~S.
andD. DeMille.
\newblock \emph{Phys. Rev. Lett.}, 100:\penalty0 203201, 2008.

\bibitem[Ichihara and Yokoyama(1995)]{IY:JCP95} A.~Ichihara
    and K.~Yokoyama.
\newblock \emph{{J. Chem. Phys.}}, 103:\penalty0 2109, 1995.

\bibitem[Idziaszek and Julienne(2010)]{Jach:03}
    Z.~Idziaszek and P.~S. Julienne.
\newblock \emph{Phys. Rev. Lett.}, 104:\penalty0 113202, 2010.

\bibitem[Idziaszek et~al.(2011)Idziaszek, Simoni, Calarco,
    and
  Julienne]{Simoni}
Z.~Idziaszek, A.~Simoni, T.~Calarco, and P.~S. Julienne.
\newblock \emph{New J. Phys.}, 13:\penalty0 083005, 2011.

\bibitem[Jambrina et~al.(2009)Jambrina, Aoiz, Eyles,
    Herrero, and {Sáez
  Rábanos}]{JAEHS:JCP09}
P.~G. Jambrina, F.~J. Aoiz, C.~J. Eyles, V.~J. Herrero, and
V.~{Sáez Rábanos}.
\newblock \emph{{J. Chem. Phys.}}, 130:\penalty0 184303, 2009.

\bibitem[Jambrina et~al.(2010{\natexlab{a}})Jambrina,
    {Alvari\~no}, Aoiz,
  Herrero, and {S\'aez R\'abanos}]{JAAHS:PCCP10}
P.~G. Jambrina, J.~M. {Alvari\~no}, F.~J. Aoiz, V.~J.
Herrero, and V.~{S\'aez
  R\'abanos}.
\newblock \emph{{Phys. Chem. Chem. Phys.}}, 12:\penalty0 12591,
  2010{\natexlab{a}}.

\bibitem[Jambrina et~al.(2010{\natexlab{b}})Jambrina, Aoiz,
    Bulut, Smith,
  {Balint-Kurti}, and Hankel]{JABSBH:PCCP10}
P.~G. Jambrina, F.~J. Aoiz, N.~Bulut, S.~C. Smith, G.~G.
{Balint-Kurti}, and
  M.~Hankel.
\newblock \emph{{Phys. Chem. Chem. Phys.}}, 12:\penalty0 1102,
  2010{\natexlab{b}}.

\bibitem[Jim\'enez-Redondo et~al.(2011)Jim\'enez-Redondo,
    Carrasco, Herrero,
  and Tanarro]{JCHT:PCCP11}
M.~Jim\'enez-Redondo, E.~Carrasco, V.~J. Herrero, and
I.~Tanarro.
\newblock \emph{{Phys. Chem. Chem. Phys.}}, 13:\penalty0 9655, 2011.

\bibitem[Julienne(2009)]{Fara} P.~Julienne.
\newblock \emph{Faraday Discuss.}, 142:\penalty0 361, 2009.

\bibitem[K.~Jachymski and Idziaszek(2013)]{Jach:02}
    P.~S.~J. K.~Jachymski, M.~Krych and Z.~Idziaszek.
\newblock \emph{Phys. Rev. Lett.}, 110:\penalty0 213202, 2013.

\bibitem[Kamisaka et~al.({2002})Kamisaka, Bian, Nobusada,
    and
  Nakamura]{KBNN:JCP02}
H.~Kamisaka, W.~Bian, K.~Nobusada, and H.~Nakamura.
\newblock \emph{{J. Chem. Phys.}}, {116}:\penalty0 {654}, {2002}.

\bibitem[Knoop et~al.(2014)Knoop, Zuchowski, Kedziera,
    Mentel, Puchalski,
  Mishra, Flores, and Vassen]{Knoop}
S.~Knoop, P.~S. Zuchowski, D.~Kedziera, A.~Mentel,
M.~Puchalski, H.~P. Mishra,
  A.~S. Flores, and W.~Vassen, 2014.
\newblock arXiv:1404.4826.

\bibitem[Kolos and Wolniewicz(1967)]{alpha} W.~Kolos and
    L.~Wolniewicz.
\newblock \emph{J. Chem. Phys.}, 46:\penalty0 1426, 1967.

\bibitem[Kotochigova(2010)]{Koto2012} S.~Kotochigova.
\newblock \emph{New Journal of Physics}, 12\penalty0 (7):\penalty0 073041,
  2010.

\bibitem[Krenos et~al.(1974)Krenos, Preston, Wolfgang, and
    Tully]{KPWT:JCP74} J.~R. Krenos, R.~K. Preston,
    R.~Wolfgang, and J.~C. Tully.
\newblock \emph{{J. Chem. Phys.}}, 60:\penalty0 1634, 1974.

\bibitem[Kutzelnigg and Jaquet(2006)]{KJ06} W.~Kutzelnigg
    and R.~Jaquet.
\newblock \emph{"Phil. Trans. R. Soc. A"}, 364:\penalty0 2855, 2006.

\bibitem[Lara et~al.(2011)Lara, Dayou, and Launay]{Lara2}
    M.~Lara, F.~Dayou, and J.-M. Launay, 2011.

\bibitem[Launay and Dourneuf(1990)]{Launayfirst} J.~M.
    Launay and M.~L. Dourneuf.
\newblock \emph{Chem. Phys. Lett}, 169:\penalty0 473, 1990.

\bibitem[Lu et~al.(2005)Lu, Chu, and Han]{LCH:JPCA05} R.~F.
    Lu, T.~S. Chu, and K.~L. Han.
\newblock \emph{{J. Phys. Chem. A}}, 109:\penalty0 6683, 2005.

\bibitem[McCall et~al.(1999)McCall, Geballe, Hinkle, and
    Oka]{MGHO:AJ99} B.~J. McCall, T.~R. Geballe, K.~H.
    Hinkle, and T.~Oka.
\newblock \emph{{Astrophys J.}}, 552:\penalty0 338, 1999.

\bibitem[M\'endez et~al.(2006)M\'endez, Gordillo, Herrero,
    and
  Tanarro]{MGHT:JPCA06}
I.~M\'endez, F.~J. Gordillo, V.~J. Herrero, and I.~Tanarro.
\newblock \emph{{J. Phys. Chem. A}}, 110:\penalty0 6060, 2006.

\bibitem[Mies(1984)]{Mies1} F.~H. Mies.
\newblock \emph{J. Chem. Phys.}, 80:\penalty0 2514, 1984.

\bibitem[Mies and Julienne(1984)]{Mies2} F.~H. Mies and
    P.~S. Julienne.
\newblock \emph{J. Chem. Phys.}, 80:\penalty0 2526, 1984.

\bibitem[Müller(1983)]{M:Diplom} D.~Müller.
\newblock \emph{Diplom thesis, University of Freiburg, Germany}, 1983.

\bibitem[Mukaiyama et~al.(2004)Mukaiyama, Abo-Shaeer, Xu,
    Chin, and
  Ketterle]{Mukaiyama}
T.~Mukaiyama, J.~R. Abo-Shaeer, K.~Xu, J.~K. Chin, and
W.~Ketterle.
\newblock \emph{Phys. Rev. Lett.}, 92:\penalty0 180402, 2004.

\bibitem[Narevicius and Raizen(2012)]{Nar3} E.~Narevicius
    and M.~G. Raizen.
\newblock \emph{Chem. Rev.}, 112:\penalty0 4879, 2012.

\bibitem[Ochs and Teloy(1974)]{OT:JCP74} G.~Ochs and
    E.~Teloy.
\newblock \emph{{J. Chem. Phys.}}, 61:\penalty0 4930, 1974.

\bibitem[Ospelkaus et~al.(2010)Ospelkaus, Ni, Wang,
    de~Miranda, Neyenhuis,
  Qu{\'{e}}m{\'{e}}ner, Julienne, Bohn, Jin, and Ye]{Ospel}
S.~Ospelkaus, K.-K. Ni, D.~Wang, M.~H.~G. de~Miranda,
B.~Neyenhuis,
  G.~Qu{\'{e}}m{\'{e}}ner, P.~S. Julienne, J.~L. Bohn, D.~S. Jin, and J.~Ye.
\newblock \emph{Science}, 327:\penalty0 853, 2010.

\bibitem[Przybytek and Jeziorski(2005)]{Przy} M.~Przybytek
    and B.~Jeziorski.
\newblock \emph{J. Chem. Phys.}, 123:\penalty0 134315, 2005.

\bibitem[Qu{\'{e}}m{\'{e}}ner
    et~al.(2004)Qu{\'{e}}m{\'{e}}ner, Honvault, and
  Launay]{Quem04}
G.~Qu{\'{e}}m{\'{e}}ner, P.~Honvault, and J.-M. Launay.
\newblock \emph{Eur. Phys. J. D.}, 30:\penalty0 201, 2004.

\bibitem[Rackham et~al.(2001)Rackham, Huarte-Larranaga, and
  Manolopoulos]{Rack0}
E.~J. Rackham, F.~Huarte-Larranaga, and D.~E. Manolopoulos.
\newblock \emph{Chem. Phys. Lett.}, 343:\penalty0 356, 2001.

\bibitem[Rackham et~al.(2003)Rackham, Gonz\'{a}lez-Lezana,
    and
  Manolopoulos]{Rack1}
E.~J. Rackham, T.~Gonz\'{a}lez-Lezana, and D.~E.
Manolopoulos.
\newblock \emph{J. Chem. Phys.}, 119:\penalty0 12895, 2003.

\bibitem[Rowe and Marquette({1987})]{RM:IJMSIP87} R.~B.
    Rowe and J.~B. Marquette.
\newblock {}.
\newblock \emph{{Int. J. Mass Spectrom. Ion. Process.}}, {80}:\penalty0 {239},
  {1987}.

\bibitem[Sadeghpour et~al.(2000)Sadeghpour, Bohn,
    Cavagnero, Esry, Fabrikant,
  Macek, and Rau]{Sade}
H.~R. Sadeghpour, J.~L. Bohn, M.~J. Cavagnero, B.~D. Esry,
I.~I. Fabrikant,
  J.~H. Macek, and A.~R.~P. Rau.
\newblock \emph{J. Phys. B: At. Mol. Opt. Phys.}, 33, 2000.

\bibitem[Schlier and Vix(1987)]{SV:CP87} C.~Schlier and
    U.~Vix.
\newblock \emph{{Chem. Phys.}}, 113:\penalty0 211, 1987.

\bibitem[Shagam and Narevicius(2012)]{Nar4} Y.~Shagam and
    E.~Narevicius.
\newblock \emph{Phys. Rev. A}, 85:\penalty0 053406, 2012.

\bibitem[Shagam and Narevicius(2013)]{Nar2} Y.~Shagam and
    E.~Narevicius.
\newblock \emph{J. Phys. Chem. C}, 117:\penalty0 22454, 2013.

\bibitem[Smith and Rowe(2000)]{SR:ACR00} I.~W.~M. Smith and
    R.~B. Rowe.
\newblock \emph{{Acc. Chem. Res.}}, 33:\penalty0 261, 2000.

\bibitem[Smith(1998)]{S:IJPC98} M.~A. Smith.
\newblock \emph{{Int. Rev. Phys. Chem.}}, 17:\penalty0 35, 1998.

\bibitem[Smith(1992)]{S:CR92} S.~D. Smith.
\newblock \emph{{Chem. Rev.}}, 92:\penalty0 1473, 1992.

\bibitem[Snow and Bierbaum(2008)]{SB:ARAC08} T.~P. Snow and
    V.~M. Bierbaum.
\newblock \emph{Ann. Rev. Anal. Chem.}, 1:\penalty0 229, 2008.

\bibitem[Sold{\'{a}}n and \v{S}pirko(2007)]{Soldan}
    P.~Sold{\'{a}}n and V.~\v{S}pirko.
\newblock \emph{J. Chem. Phys.}, 127:\penalty0 121101, 2007.

\bibitem[Sold\'{a}n et~al.(2002)Sold\'{a}n, Cvita\v{s},
    Hutson, Honvault, and
  Launay]{Sol02}
P.~Sold\'{a}n, M.~T. Cvita\v{s}, J.~M. Hutson, P.~Honvault,
and J.-M. Launay.
\newblock \emph{Phys. Rev. Lett.}, 89:\penalty0 153201, 2002.

\bibitem[Staanum et~al.(2006)Staanum, Kraft, Lange, Wester,
    and
  {Weidem\"{u}ller}]{Staanum}
P.~Staanum, S.~D. Kraft, J.~Lange, R.~Wester, and
M.~{Weidem\"{u}ller}.
\newblock \emph{Phys. Rev. Lett.}, 96:\penalty0 023201, 2006.

\bibitem[Syassen et~al.(2006)Syassen, Volz, Teichmann,
    {D\"{u}rr}, and
  Rempe]{Syassen}
N.~Syassen, T.~Volz, S.~Teichmann, S.~{D\"{u}rr}, and
G.~Rempe.
\newblock \emph{Phys. Rev. A}, 74:\penalty0 062706, 2006.

\bibitem[Takayanagi et~al.(2000)Takayanagi, Kurosaki, and
    Ichihara]{TKI:JCP00} T.~Takayanagi, Y.~Kurosaki, and
    A.~Ichihara.
\newblock \emph{{J. Chem. Phys.}}, 112:\penalty0 2615, 2000.

\bibitem[Teloy and Gerlich({1974})]{TG:CP74} E.~Teloy and
    D.~Gerlich.
\newblock \emph{{Chem. Phys.}}, {4}:\penalty0 417, {1974}.

\bibitem[T.J.Millar(2005)]{M:AG05} T.J.Millar.
\newblock \emph{{Astrophys and Geophys.}}, 46:\penalty0 2.29, 2005.

\bibitem[Tosi({1992})]{T:CR92} P.~Tosi.
\newblock \emph{{Chem. Rev.}}, {92}:\penalty0 1667, {1992}.

\bibitem[van~de Meerakker and Meijer(2009)]{FaraBas}
    S.~van~de Meerakker and G.~Meijer.
\newblock \emph{Faraday Discuss.}, 142:\penalty0 113, 2009.

\bibitem[Vassen et~al.(2012)Vassen, Cohen-Tannoudji, Leduc,
    Boiron, Westbrook,
  Truscott, Baldwin, Birkl, Cancio, and Trippenbach]{Vassen}
W.~Vassen, C.~Cohen-Tannoudji, M.~Leduc, D.~Boiron, C.~I.
Westbrook,
  A.~Truscott, K.~Baldwin, G.~Birkl, P.~Cancio, and M.~Trippenbach.
\newblock \emph{Rev. Mod. Phys.}, 84:\penalty0 175, 2012.

\bibitem[Velilla et~al.(2008)Velilla, Lepetit, Aguado,
    Beswick, and
  Paniagua]{VLABP:JCP08}
L.~Velilla, B.~Lepetit, A.~Aguado, J.~A. Beswick, and
M.~Paniagua.
\newblock \emph{{J. Chem. Phys.}}, 129:\penalty0 084307, 2008.

\bibitem[Viegas et~al.(2007)Viegas, Alijah, and
    Varandas]{VAV:JCP07} L.~P. Viegas, A.~Alijah, and
    A.~J.~C. Varandas.
\newblock \emph{{J. Chem. Phys.}}, 126:\penalty0 074309, 2007.

\bibitem[Villinger et~al.(1982)Villinger, Henchman, and
    Lindinger]{VHL:JCP82} H.~Villinger, M.~J. Henchman, and
    W.~Lindinger.
\newblock \emph{{J. Chem. Phys.}}, 76:\penalty0 1590, 1982.

\bibitem[Watson(1973)]{W:AJ73a} W.~D. Watson.
\newblock \emph{{Astrophys J.}}, 181:\penalty0 L129, 1973.

\bibitem[Weiner et~al.(1999)Weiner, Bagnato, Zilio, and
    Julienne]{Weiner} J.~Weiner, V.~S. Bagnato, S.~Zilio,
    and P.~S. Julienne.
\newblock \emph{Rev. Mod. Phys.}, 71:\penalty0 1, 1999.

\bibitem[Wynar et~al.(2000)Wynar, Freeland, Han, Ryu, and
    Heinzen]{Wynar} R.~Wynar, R.~S. Freeland, D.~J. Han,
    C.~Ryu, and D.~J. Heinzen.
\newblock \emph{Science}, 287:\penalty0 1016, 2000.

\bibitem[Zahzam et~al.(2006)Zahzam, Vogt, Mudrich,
    Comparat, and
  Pillet]{Zahzam}
N.~Zahzam, T.~Vogt, M.~Mudrich, D.~Comparat, and P.~Pillet.
\newblock \emph{Phys. Rev. Lett.}, 96:\penalty0 023202, 2006.

\bibitem[Zanchet et~al.(2009)Zanchet, Roncero,
    Gonz\'alez-Lezana,
  Rod\'{\i}guez-L\'opez, Aguado, Sanz-Sanz, and
  {G\'omez--Carrasco}]{ZRGRASG:JPCA09}
A.~Zanchet, O.~Roncero, T.~Gonz\'alez-Lezana,
A.~Rod\'{\i}guez-L\'opez,
  A.~Aguado, C.~Sanz-Sanz, and S.~{G\'omez--Carrasco}.
\newblock \emph{{J. Phys. Chem. A}}, 113:\penalty0 14488, 2009.

\bibitem[Zhang et~al.(2009)Zhang, Zhang, and
    Chen]{ZZC:JTCC09} C.~H. Zhang, W.~Q. Zhang, and M.~D.
    Chen.
\newblock \emph{{J. Theor. Comp. Chem.}}, 8:\penalty0 403, 2009.

\bibitem[Zhang et~al.(2010)Zhang, Li, Chu, and
    Chen]{ZLXC:CP10} W.~Zhang, Y.~Li, X.~Chu, and M.~D.
    Chen.
\newblock \emph{{Chem. Phys.}}, 367:\penalty0 115, 2010.

\bibitem[Zhang and Chen(2009)]{ZC:JTCC09} W.~Q. Zhang and
    M.~D. Chen.
\newblock \emph{{J. Theor. Comp. Chem.}}, 8:\penalty0 1131, 2009.

\bibitem[Zieger et~al.(2010)Zieger, van~de Meerakker,
    Heiner, Bethlem, van
  Roij, and Meijer]{Syncro}
P.~C. Zieger, S.~Y.~T. van~de Meerakker, C.~E. Heiner,
H.~L. Bethlem, A.~J.~A.
  van Roij, and G.~Meijer.
\newblock \emph{Phys. Rev. Lett.}, 105:\penalty0 173001, 2010.

\end{thebibliography}

\end{document}